\documentclass[a4paper,fleqn,usenatbib]{mnras}
\usepackage{mathptmx}
\usepackage[T1]{fontenc}
\usepackage{ae,aecompl}

\usepackage{graphicx}	% Including figure files
\usepackage{amsmath}	% Advanced maths commands
\usepackage{amssymb}	% Extra maths symbols
\usepackage{float}
\usepackage{textcomp}
\usepackage{hyperref}
\hypersetup{colorlinks=true,linkcolor=red,urlcolor=cyan,citecolor=blue}
\usepackage{multicol}
\usepackage{subfig}
\usepackage{float, caption}
\usepackage{lscape}
\title[Systematic errors in SL modelling]{Hubble Frontier Fields: systematic errors in strong lensing models of galaxy clusters - Implications for cosmography}
\author[A. Acebron et al.]{
	\parbox{\textwidth}{
Ana Acebron$^{1}$\thanks{E-mail: ana.acebron@lam.fr},
Eric Jullo$^{1}$,
Marceau Limousin$^{1}$,
Andr\'e Tilquin$^{2}$, 
Carlo Giocoli$^{1,3}$,
Mathilde Jauzac$^{4,5,6}$,
Guillaume Mahler$^{7}$
and Johan Richard$^{7}$}
\\
\\
$^{1}$ Aix Marseille Univ, CNRS, LAM, Laboratoire d'Astrophysique de Marseille, Marseille, France \\
$^{2}$ Aix Marseille Univ., CNRS, CPPM, Marseille, 13288 Marseille cedex 09, France\\
$^{3}$ Dipartimento di Fisica e Astronomia, Alma Mater Studiorum Universit\`{a} di Bologna, via Gobetti 93/2, 40129, Bologna, Italy \\
$^{4}$ Centre for Extragalactic Astronomy, Department of Physics, Durham University, Durham DH1 3LE, U.K. \\
$^{5}$ Institute for Computational Cosmology, Departement of Physics, University of Durham, South Road, Durham DH1 3LE, U.K\\
$^{6}$ Astrophysics and Cosmology Research Unit, School of Mathematical Sciences, University of KwaZulu-Natal, Durban 4041, South Africa\\
$^{7}$ Univ Lyon, Univ Lyon1, Ens de Lyon, CNRS, Centre de Recherche Astrophysique de Lyon UMR5574, F-69230, Saint-Genis-Laval, France
}

\date{Accepted XXX. Received YYY; in original form ZZZ}

\pubyear{2017}

\begin{document}
\label{firstpage}
\pagerange{\pageref{firstpage}--\pageref{lastpage}}
\maketitle

\begin{abstract}
  {Strong gravitational lensing by galaxy clusters is a fundamental tool to study dark matter and constrain the geometry of the Universe. Recently, the HST Frontier Fields program has allowed a significant improvement of mass and magnification measurements but lensing models still have a residual RMS between 0.2" and few arcseconds, not yet completely understood. Systematic errors have to be better understood and treated in order to use strong lensing clusters as reliable cosmological probes.
  We have analysed two simulated Hubble Frontier Fields-like clusters from the Hubble Frontier Fields Comparison Challenge, \textit{Ares} and \textit{Hera}. We use several estimators (relative bias on magnification, density profiles, ellipticity and orientation) to quantify the goodness of our reconstructions by comparing our multiple models, optimized with the parametric software \textsc{Lenstool}, with the input models. We have quantified the impact of systematic errors arising, first, from the choice of different density profiles and configurations and, secondly, from the availability of constraints (spectroscopic or photometric redshifts, redshift ranges of the background sources) in the parametric modelling of strong lensing galaxy clusters and therefore on the retrieval of cosmological parameters.
  We find that substructures in the outskirts have a significant impact on the position of the multiple images, yielding tighter cosmological contours. The need for wide-field imaging around massive clusters is thus reinforced. We show that competitive cosmological constraints can be obtained also with complex multimodal clusters and that photometric redshifts improve the constraints on cosmological parameters when considering a narrow range of (spectroscopic) redshifts for the sources.}

\end{abstract}


%%%%%%%%%%%%%%%%%%%%%%%%%%%%%%%%%%%%%%%%%%%%%%%%%%

%%%%%%%%%%%%%%%%% BODY OF PAPER %%%%%%%%%%%%%%%%%%

\section{Introduction}
Strong gravitational Lensing (SL, hereafter) is nowadays at the very heart of important issues in modern cosmology, allowing for instance to measure directly the total projected mass distribution (baryonic and dark) of galaxy cluster cores \citep[e.g.][]{Richard2010, Zitrin2016, Monna2017, Mahler2017}, to image very high redshift sources otherwise too faint to be detected without the gravitational magnification \citep{kneib2004, richard2008, Coe2013, Atek2015} and to constrain the geometry of the Universe \citep{jullo2010, Daloisio2011, Magana2015, Caminha2016} which is one of the long-standing challenges of modern cosmology. \\
In the standard cosmological model $\Lambda\mathrm{CDM}$, $\sim$ 72$\%$ of the energy density of the Universe is in the form of a 'dark energy', a fluid with negative pressure that would cause the presently acceleration of the Universe \citep{Planck2016}.\\
Type Ia supernovae \citep{Riess1998}, baryon acoustic oscillations \citep{Beutler2011, Martin2016}, cosmic shear \citep{Massey2007, Heymans2012, Hildebrandt2017}, cluster abundances \citep{deHaan2016}, CMB anisotropies \citep{Planck2016} or time delays \citep{Suyu2016} are some of the several probes allowing to better understand the constituents of the Universe and their properties by putting tighter constraints on cosmological parameters. \\
However, in order to obtain robust estimates of cosmological parameters, estimates from the  different cosmological probes must be combined as each technique has distinct degeneracies and biases \citep{Planck2016b, Peel2017}.\\
Among the previously mentioned cosmological probes, using the strong lensing features in galaxy clusters is a very promising technique that yields orthogonal constraints in an era of precise cosmology \citep{golse2002, gilmore2009, jullo2010}. To perform cosmography with strong lenses, a precise and accurate mass distribution of the cluster is required, i.e, a large number of constraints is crucial.\\
Recently, the \textit{Hubble Frontier Fields} program\footnote{\url{http://www.stsci.edu/hst/campaigns/frontier-fields}} \citep[HFF; P.I.: J. Lotz,][]{Lotz2017} with the \textit{Hubble Space Telescope} (hereafter HST) has provided the deepest multi-colour imaging of galaxy clusters to date, that combined with spectroscopy from ground-based surveys, has led to the discovery of hundreds of multiple images and thus to a significant improvement of cluster mass estimates \citep{Jauzac2014, Diego2016, Lagattuta2016, Monna2017}.\\
The mass modelling of strong lensing clusters can be carried out in different manners: parametric and non-parametric methods are equally used; the primary distinction between them being that parametric modelling assumes that luminous cluster galaxies trace the cluster mass whereas non-parametric does not.\\
The FF-SIMS Challenge \citep{meneghetti2016}, an archive of mock HFF-like clusters, has provided the lensing community with a set of simulated clusters in order to highlight for the first time the strengths and weaknesses of each methods: parametric and non-parametric. This work would allow the non-lensing community to choose a method and software according to their different needs. This challenge has shown that all lensing reconstruction methods provide reliable mass distributions.
However, strong lensing modelling appears to be still unable to match the HST observations angular resolution ($\sim$  0.05") with a residual RMS between 0.2" and a few arcseconds \citep{limousin2016}, a systematical error not yet completely understood and few studies have addressed this issue. \\
Indeed, strong lensing mass modelling has various sources of systematic errors, arising from the hypothesis behind our models, which have recently started to be acknowledged and analysed in a more quantitative way. \citet{Meneghetti2010b} studied the properties of $\sim50 000$ strong lensing clusters in the MARENOSTRUM cosmological simulation. They find that strong lenses tend to have their major axes oriented along the line of sight. This orientation bias results for instance in cluster concentrations estimations from the projected density profiles to be biased high \citep{Giocoli2014, Sereno2015}. \citet{Daloisio2011} quantified  that the modelling errors due to the scatter in the cluster-galaxy scaling relations and unmodelled line-of-sight haloes can result in errors of the order of a few arcseconds on average. \citet{Bayliss2015} studied the impact of assuming a certain cosmological model on the determination of magnification and mass profiles of the cluster core for HFF clusters showing that cosmological parameter uncertainty is a non-negligible source of errors for the lens modelling.\\ 
Foreground or background large-scale structures impacts the SL modelling as well \citep{Dalal2005,Host2012}, that could introduce a systematic error of up to $\sim 1.4"$ on the position of multiple images \citep{Zitrin2015}. Line of sight effects have then to be taken into account during the strong lensing modelling in order to recover precise cosmological parameters \citep{jullo2010, Caminha2016}.\\
Systematic errors can also arise from the mass distributions assumed for the dark matter components as shown in \citet{limousin2016} where the observed constraints in MACS0717 are equally well reproduced by a mass model with a shallow large-scale DM component and one for which this component is peaky. \citet{Harvey2016} analysed the FF cluster MACSJ0416 ($z=0.397$) and found that the assumption that light traces mass can introduce an error of $\sim$0.5" on the position of the multiple images. \citet{Bouwens2016} studied the impact of magnification uncertainties on luminosity functions from the first four HFF clusters. \citet{Johnson2016} have led the first investigation attempting to quantify systematic errors induced by the availability of constraints in strong lensing clusters. They show that the accuracy of the magnification is sensitive to the selection of constraints rather than their amount. \\ 
In this paper we take advantage of the mock HFF-like clusters archive from the FF-SIMS Challenge and use two of the HFF-like mock strong lensing clusters (\textit{Ares} and \textit{Hera}) to investigate how systematic errors in the strong lensing parametric modelling with \textsc{Lenstool} affect the determination of the total mass distribution in clusters and hence the retrieval of robust cosmological parameters such as the mean matter density and dark energy equation of state parameters $\mathrm{\Omega_{M}}$ and $w$, respectively.\\
First, we use four different estimators (density profiles, relative bias on magnification, cluster's ellipticity and orientation angle) to compare our reconstructions with the input models in order to assess the impact of different mass distributions and configurations in the strong lensing modelling of clusters and hence on the cosmography. \\ 
Secondly, we study how the availability of constraints affects the retrieval of unbiased cosmological parameters. We investigate if there is a more efficient range of redshifts to recover the input cosmology, if including photometric redshifts can result in a more robust estimation of these parameters and if taking into account an increasing number of photometric families translates into an increasing precision in the cosmological parameters estimation.\\
The paper is organized as follows: the data used for this analysis is presented in Section \ref{[section2]}, in Section \ref{[section3]} we describe the methodology, in Section \ref{[SLM]} we detail the different modellings for \textit{Ares} and \textit{Hera}, we then show the systematic uncertainties in Section \ref{[results]}, their impact on cosmography in Section \ref{cosmosec} and conclude in Section \ref{[conclu]}.\\
Throughout the paper, we use the standard ${\Lambda\mathrm{CDM}}$  flat cosmological model with the Hubble constant $H_0 = 70 \mathrm{km s^{-1} Mpc^{-1}}$, $\mathrm{\Omega_{M}}$ and $w$ are let as free parameters. Magnitudes are quoted in the AB system.

\begin{figure*}
	\centering 
	\includegraphics[width=1.33\columnwidth]{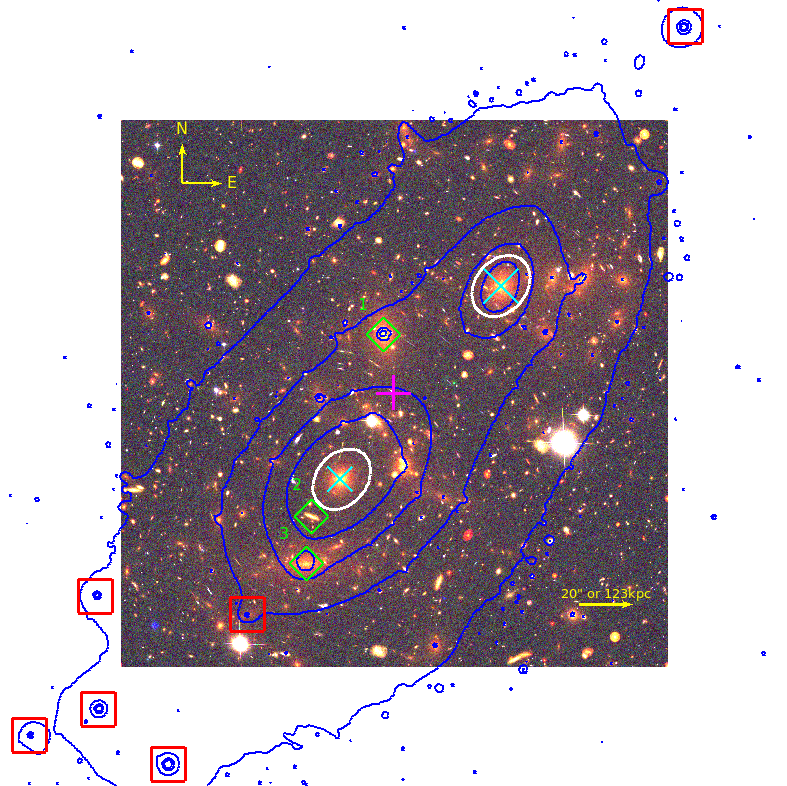}
	\includegraphics[width=1.33\columnwidth]{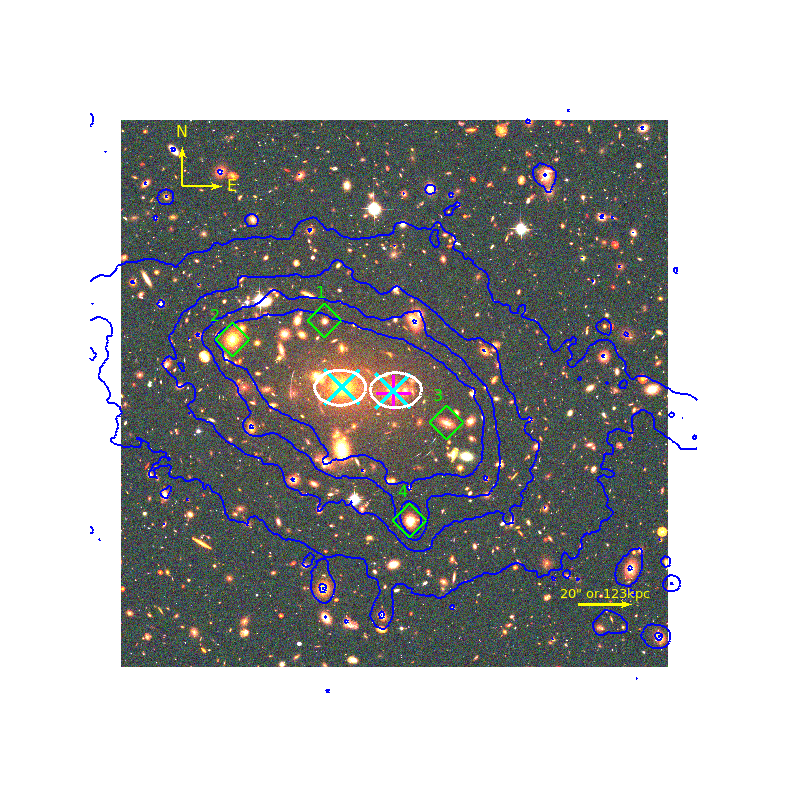}
	\caption{The underlying colour image is a composite created from the provided HST/ACS simulated images in the F814W, F606W, and F435W passbands. Input convergence map are shown in blue.\\ Upper panel: \textit{Ares} cluster. The cluster's centre is indicated by the magenta cross. Large-scale potentials are shown as white ellipses, BCGs as cyan crosses, galaxy haloes as green diamonds and red boxes for the outskirts substructures.\\ Bottom panel: \textit{Hera} cluster. Same colour code as for \textit{Ares}.}
	\label{clusters}
\end{figure*}

\section{The Data}\label{[section2]} %
We have analysed two mock HFF-like clusters from the FF-simulations Challenge archive \citep{meneghetti2016}:
\textit{Ares} and \textit{Hera}. They were initially created for the lens modelling comparison project, which, for the first time, compares the reconstructions obtained using different techniques, such as parametric, non-parametric and hybrid, performed by different teams. \\
Both clusters have been simulated to reproduce, not only the characteristics of the \textit{HST Advandced Camera for Survey} (ACS) and \textit{Wide-Field Camera 3} (WFC3) observations (depth, passbands and spatial resolution), but also the complexity of the FF clusters themselves even if these two have been created using different techniques. \\
Both clusters have then been modelled to be realistic, complex, bi-modal clusters (see Figure \ref{clusters}), generated in a flat $\mathrm{\Lambda_{CDM}}$  cosmological model.\\
During the challenge, all participants first performed a blind analysis (i.e, not knowing the true mass distribution) of these clusters for which only the HFF-like images as well as catalogues of multiple images (position and spectroscopic redshifts) and cluster galaxies (positions and magnitudes in all ACS/WFC3 bands) were provided with $m_{F814W} < 24$ (see Table \ref{table:1} for further details). \\
\textit{Ares} is a more powerful lens than \textit{Hera}, producing thus many more multiple images \citep{meneghetti2016}. Their respective redshift distributions are shown in Figure \ref{z_areshera}. The source galaxies resemble the luminosity and the redshift distribution of the galaxies in the \textit{Hubble Ultra-Deep-Field} \citep{Coe2006}.\\
For both clusters, the software \textsc{Skylens} \citep{meneghetti2008} was used to ray-trace the lensed galaxies to the image plane and to create simulated HFF-like images in all bands. \\
After the unblinding, the convergence and shear maps (calculated for a source at $z_{s}$=9) were provided. They were used to compare our reconstructions with the true values and also to improve our models. The input cosmology was also revealed.
We refer the reader to \citet{meneghetti2016} for a detailed presentation on how \textit{Ares} and \textit{Hera} mass distributions were generated, here we only present a quick overview.\\
\subsection{\textit{Ares}}\label{aresdata} 
\textit{Ares} is a semi-analytical cluster \citep[created using MOKA\footnote{\url{https://cgiocoli.wordpress.com/research-interests/moka/}} by][]{giocoli2012} at z=0.5. This simulated cluster is built with three components: two smooth dark matter triaxial haloes with a NFW profile, a bright central cluster galaxy (BCG) with an Hernquist profile \citep{hernquist1990} and sub-haloes having a Singular Isothermal Sphere profile \citep{hernquist1990}. Dark matter sub-haloes are populated using a Halo Occupation Distribution technique (HOD) and stellar and B-band luminosities are given for all galaxies according to the mass of their sub-halo as in \citet{wang2006}.
The mass within the virial radius is then defined as $M_{vir} = M_{smooth} + \Sigma_{i} m_{sub,i}$.\\
\textit{Ares} is generated in a flat ${\Lambda\mathrm{CDM}}$  cosmological model with a matter density parameter  $\mathrm{\mathrm{\Omega_{M}}}=0.272$.
\subsection{\textit{Hera}}\label{heradata}
\textit{Hera} is from a N-body simulation of cluster-sized dark matter halos for a flat ${\Lambda\mathrm{CDM}}$  model \citep[see][]{Planelles2014} at z=0.507 which was re-simulated using a TreePM-SPH GADGET-3 code with only collision-less dark matter particles. The properties of cluster galaxies are created from Semi-Analytic Methods of galaxy formation \citep{DeLucia2007}.
\textit{Hera} is also created in a flat ${\Lambda\mathrm{CDM}}$  cosmological model with a matter density parameter  $\mathrm{\Omega_{M}}=0.24$.\\

\begin{table}
	\caption{Further details for \textit{Ares} and \textit{Hera} clusters}             % title of Table
	\label{table:1}      % is used to refer this table in the text
	\centering                          % used for centering table
	\begin{tabular}{c c c c c}        % centered columns (4 columns)
		\hline\hline                 % inserts double horizontal lines
		Cluster name & $ z $ & Cluster galaxies & Images & Sources \\    % table heading 
		\hline          % inserts singe horizontal line
		
		\\    \textit{Ares} & 0.5 & 330 & 242 & 85\\      % inserting body of the table
		\textit{Hera} & 0.507 & 337  & 65 & 19\\
		\\ 
		\hline\hline                                  %inserts single line
	\end{tabular}
\end{table}

For both clusters, the software \textsc{Skylens} \citep{meneghetti2008} was used to ray-trace the lensed galaxies to the image plane and to create simulated HFF-like images in all bands. \\

\section{Methodology} \label{[section3]}

\subsection{Mass modelling} \label{massmod}
We perform the strong lensing modelling in the source plane using the public software \textsc{Lenstool}\footnote{\url{https://projets.lam.fr/projects/lenstool}} \citep{kneib1996, jullo2007} which performed very well for the FF-SIMS Challlenge. \textsc{Lenstool} utilizes a Bayesian Markov chain Monte-Carlo \citep[hereafter MCMC, see][for details]{jullo2007} sampler to optimize the model using the positions of the multiply imaged systems. The matter distribution of clusters is decomposed into several smooth large scale components and individual contributions from cluster galaxies. In this work, the reconstructions are based on the strong lensing information only.\\

We have modelled both clusters as up to three components: i) large-scale potentials, ii) brightest cluster galaxies (BCGs) and iii) individual galaxies identified spectroscopically, with masses scaling with luminosity. Each mass component has a parameterized profile such as the Pseudo Isothermal Elliptical Mass Distribution \citep[hereafter PIEMD]{kassiola1993}, Navarro, Frenk \& White (NFW) \citep{Navarro1997} or Hernsquist \citep{hernquist1990}. For instance, large-scale potentials are modelled with either a NFW or PIEMD, BCGs with PIEMD or Hernquist profiles and individual clusters members with a PIEMD profile, scaled according to the relations in \citet{limousin2005}. These mass distributions are briefly presented hereafter.\\

We used the NFW density profile \citep{Navarro1997} to model \textit{Ares} and \textit{Hera} large-scale mass distributions. The 3D density profile is given by:
\begin{equation}
\centering
\rho_{NFW}(r)= \dfrac{\rho_{s}}{(\dfrac{r}{{r_{s}}})(1 + \dfrac{r}{{r_{s}}})^{2}}\mathrm{,}
\end{equation}

where $\mathrm{\rho_{s}}$ is a characteristic density and $r_{s}$ is the scale radius. \\
This profile behaves as $\rho \propto r^{-1}$ in the inner region, $\rho \propto r^{-2}$ at $r=r{_s}$ and as $\rho \propto r^{-3}$ in the outer regions.\\
In \textsc{Lenstool}, this profile has the following free parameters: x and y, the
coordinates of the halo centre; e, the ellipticity, defined as e = $(a^{2} +b^{2} )/ (a^{2} - b^{2})$ with a and b the semi-major and semi-minor axis respectively and $\theta$, the position angle (counted counter-clockwise from the x axis); $r_{s}$, the scale radius and $\sigma$, the velocity dispersion. \\

The pseudo isothermal density profile \citep{limousin2005} is used to model dark matter haloes and/or individual galaxies which 3D density distribution is given by: 
\begin{equation}
\centering
\rho_{PIEMD}(r) = \dfrac{\rho_{0}}{(1 + \dfrac{r^{2}}{{r_{core}}^{2}})(1 + \dfrac{r^{2}}{{r_{cut}}^{2}})}\mathrm{,}
\end{equation}
with a core radius $r_{core}$ and a truncation radius $r_{cut}$.\\
This profile is characterized by two changes in the density slope: within the transition region ( $r_{core}$ < r < $r_{cut}$) it behaves as an isothermal profile with $\rho \propto r^{-2}$, while exiting this region, the density will fall as $\rho \propto r^{-4}$ (such behaviour is common for elliptical galaxies). \\
It has the following free parameters: the coordinates x, y; the ellipticity, e; angle position, $\theta$; core and cut radii, $r_{core}$ and $r_{cut}$ and a velocity dispersion, $\sigma$.\\

We use the Hernquist profile \citep{hernquist1990} to model the BCGs of the two clusters. BCGs are massive elliptical galaxies with observed luminosities well represented by the de Vaucouleurs $R^{\dfrac{1}{4}}$ empirical law (which fits well observations) but being more simple analytically. The 3D density profile of te Hernquist profile can be written as follows:
\begin{equation}
\centering
\rho_{H}(r) = \dfrac{M}{2\mathrm{\pi}}\dfrac{a}{r}\dfrac{1}{(r + a)^{3}}\mathrm{,}
\end{equation}
where $M$ is the total mass and $a$, the characteristic scale length. This profile is extremely similar to the NFW profile at small radii but the density falls as $\rho \propto r^{-4}$ at larger radii. \\
It is parametrized using the following free parameters: the coordinates x, y; the ellipticity, e; angle position, $\theta$; a core radius, $r_{core}$ and a velocity dispersion, $\sigma$. \\

Once the mass components are defined, the best-fitting model parameters are found by minimizing the distance between the observed and model-predicted positions of the multiple images, and the parameter covariance is estimated using a Bayesian Markov Chain Monte Carlo (MCMC) technique \citep{jullo2007}. For each of the models, we use several statistical values to assess the goodness of the fit and to discriminate between models.
\begin{enumerate}
	\item We use the root-mean-square between the observed and predicted positions of the multiple images from the modelling, computed as follows: 
	\begin{equation}
	RMS = \sqrt{\dfrac{1}{N} \sum_{i=1}^{n} |\theta_i^{obs} - \theta_i^{pred}|^2}\mathrm{,}
	\end{equation}
	where $\theta_i^{obs}$ and $\theta_i^{pred}$ are the observed and model-predicted positions of the multiples images and N being the total number of images.
	\item We compute the Bayesian Information Criterion \citep[BIC, introduced by][]{Schwarz1978}:
	\begin{equation}
	BIC = -2\ln(L) +k \times \ln(n)\mathrm{,}
	\end{equation}
	with $L$, the likelihood; $k$, the number of free parameters and $n$, the number of constraints.
    \item The corrected Akaike Information Criterion (AICc). The BIC will over-penalize models whereas AICc tends to under-penalize. Using the two together helps balancing those effects. It is computed as follows:
   	\begin{equation}
    AICc = 2k - \ln(L) + \frac{2k(k+1)}{(n - k -1)}\mathrm{,}
    \end{equation}
	\item The reduced $\chi^2$, see \citet{jullo2007} for details.
	\item The Bayesian Evidence which considers the 'complexity' of the models and how this 'complexity' is justified by the observables, see also \citet{jullo2007} for details.
\end{enumerate}
\subsection{Cosmological parameters} \label{cosmo}
As both \textit{Ares} and \textit{Hera} have a large number of multiple images with spectroscopic redshifts for all images -up to redshift $z_{s} \sim 6$ for \textit{Ares} and $z_{s} \sim 3.5$ for \textit{Hera} (see Figure \ref{z_areshera})- they represent good probes to constrain cosmological parameters. \\ 
We briefly outline here the methodology to estimate these parameters with cluster strong lensing \citep[further details can be found in][for instance]{golse2002, gilmore2009}. \\
Strong lensing is sensitive to the underlying geometry of the Universe as the position of the multiple images not only depends on the mass distribution of the lens, but also on the angular diameter distances from the observer to the lens ($D_{OL}$), to the source ($D_{OS}$), and from the lens to the source ($D_{LS}$). This dependence is used to put constraints on the cosmological parameters $\mathrm{\mathrm{\Omega_{M}}}$ and $w$. \\ 
Indeed, the lens equation can be written as:

\begin{equation}
\centering
\vec{\beta_{i}} = \vec{\theta_{i}} -\dfrac{2}{c^{2}} \dfrac{D_{LS}}{D_{OL}D_{OS}} \nabla \phi(\vec{\theta_{i}})\mathrm{,}
\end{equation}

where $\theta$ and $\beta$ are the (multiple) image angular positions in the lens and source planes respectively, $\phi$ is the projected Newtonian potential of the lens and the cosmological dependence is embedded into the angular diameter distances.\\
When only one family of multiple images is available, the ratio between the cosmological distances cannot be disentangled from the gradient of the potential. However, if at least 2 systems of images at different redshifts are available, this degeneracy can be broken and  via the 'family ratio' \citep[see][]{Link1998} from which constraints on $\mathrm{\Omega_{M}}$ and $w$ can be obtained:
\begin{equation}
\Xi(z_L,z_{s1},z_{s2};\Omega_M,w)= \dfrac{D_{LS_1}D_{OS_2}}{D_{OS_1}D_{LS_2}}\mathrm{,}
\end{equation}
with $\mathrm{z_L}$ is the redshift of the lens,  $z_{s1}$ and $z_{s2}$ are the redshifts of two distinct source and $D$ is the angular diameter distance.\\
This kind of analysis has already been carried out for galaxy clusters Abell 2218, Abell 1689 and Abell S1063 (or RXC J2248.7-4431) by \citet{soucail2004, jullo2010, Caminha2016} respectively and also with simulated data \citep{golse2002, gilmore2009, Daloisio2011}.\\
In this work, the energy density of the total matter of the universe $\mathrm{\Omega_{M}}$ and the equation-of-state parameter $w$ are set as free parameters in a ${\Lambda\mathrm{CDM}}$ flat cosmological model.

\subsection{Priors} \label{priors}
In this section we describe the choice of boundaries for the flat priors of the free parameters mentioned below. Some boundaries are set with a quite large interval. However others have more narrow priors.\\
The positions of the haloes (for both cluster- and galaxy- scale potentials), x and y, correspond to the light peak of galaxies (the position of the BCGs being the central coordinates for the dark matter haloes). The angle position of haloes are also set by the luminous component's angle. The ellipticity is not allowed to reach very high values following \citet{Despali2017}. This work shows that high ellipticities ($e>0.75$) are not favored by theoretical predictions.\\
Finally, some parameters have quite narrow priors as models have been run several times and in order to gain in computing time some prior ranges were tighten, checking that the boundaries were not reached.
Nonetheless, we checked that a slight change in the boundaries had no impact on the posterior distribution of the parameters. 

\section{Strong lensing models} \label{[SLM]}
We here present the details of each model for the \textit{Ares} and \textit{Hera} clusters, the free parameters for each potential and the respective flat priors.
These models are presented in chronological order as we tried to model the clusters in a more complex way (and also after the unblinding of the true mass profiles).\\
Their id number throughout the paper is the last number of their corresponding subsection. The first term stands for the density profile used for the large-scale haloes, the second (if any) for BCGs. \\
All coordinates are presented  as the distance to the centre of the clusters (in arcseconds) shown as the magenta cross in Figure \ref{clusters}. The centre has been arbitrarily chosen to be ($\delta R.A., \delta Dec$)= (0", 0") and serves as reference for \textsc{Lenstool}.

\subsection{\textit{Ares}} \label{[section4]}
Thanks to the HST simulated images, we easily see that this cluster is bimodal, all the models thereafter will have two large-scale haloes whose coordinates stated below correspond to those of the light peaks (see upper panel of Figure \ref{clusters}).
Cluster member galaxies are taken from the given simulated catalogues up to a magnitude of $m_{F160W}$ < 22.0 (being a more complex cluster, we limited the sample with a magnitude cut to gain in computing time, representing $>90\%$ of the total cluster luminosity) and the small scale haloes associated to galaxy members are parametrized with a PIEMD profile with a fixed core radius of 0.15 kpc, a velocity dispersion $\sigma^{\star}$ allowed to vary between 94 and 180 km/s, a cut radius $r_{cut}^{\star}$ varying from 78 kpc to 272.00 kpc, considered spherical, with a $mag0=18.5$ (the reference magnitude for the scaling relations corresponding to $L^{\star}$ at the cluster's redshift) and following the scaling relations \citep{Faber1976}. \\
Moreover, three massive galaxies were more carefully modelled (shown as green diamonds in Figure \ref{clusters}) also using a PIEMD density profile. These massive galactic haloes were modelled independently (not using the scaling relations) as they appeared to have a significant impact in the mass profile and/or on the nearby multiple images. For the three of them, the cut radius $r_{cut}^{\star}$ is fixed to 1000 kpc as initially the modelling was undertaken as the merger of four haloes (supported by the high velocity dispersions of these massives galaxies).\\
The first galaxy (labelled as 1 in Figure \ref{clusters}) is fixed at ($\delta R.A., \delta Dec$)= (+4.053", +22.042") away from the centre of the cluster. Its ellipticity can go up to 0.6, the angle position is allowed to vary from 60.0 to 120.0 degrees and its velocity dispersion $\sigma$ from 100.0 to 400.0 km/s.\\
The galaxy labelled as 2 in Figure \ref{clusters} is located at ($\delta R.A., \delta Dec$)= (+30.78", -45.98") away from the centre of the cluster with an ellipticity from 0.3 to 0.7, the angle position is allowed to vary from 0.0 to 90.0 degrees and its velocity dispersion $\sigma$ from 50.0 to 400.0 km/s.\\
Finally, the galaxy labelled as 3 in Figure \ref{clusters} is located at ($\delta R.A., \delta Dec$)= (+33.008", -63.542") away from the centre of the cluster. From the input convergence contours it appeared as a massive halo. It is modelled with ellipticity with values between 0.1 - 0.7, the angle position is allowed to vary from 141 to 171 degrees and its velocity dispersion $\sigma$ from 400.0 to 700.0 km/s.\\
All multiple images are included in the models with a positional uncertainty of 0.5" and the optimization is performed in the source plane as it is less computing time expensive. We checked that the results are similar with both source and image plane optimizations.  

\subsubsection{PIEMD } \label{mod1}
This first model contains two large-scale haloes whose coordinates are fixed at ($\delta R.A., \delta Dec$)= (+20.0", -32.0") and (-40.0", +40.0") away from the centre of the cluster respectively. These clumps do not harbour any additional halo linked to the BCGs. \\
For the first clump, we let the core radius vary from 20 kpc to 65 kpc with a fixed cut radius of 1000 kpc. The ellipticity of this halo is allowed to reach values from 0.2 and as high as 0.7 and its orientation from 130 to 140 degrees and the velocity dispersion can vary from 400 to 1700 km/s.
The second large-scale halo can have a core radius with any value from 20 kpc to 60 kpc and the cut radius is fixed to 1000 kpc. Its ellipticity can take any value from 0.3 to 0.6, its orientation can vary from 105 to 115 degrees and the velocity dispersion can vary from 400 to 1000 km/s.

\subsubsection{PIEMD - PIEMD }\label{mod2}
This reconstruction is the same as the previous one but we add two BCG components (modelled using the PIEMD profile) whose coordinates are fixed to main large-scale haloes'. These BCGs are modelled separately and do not follow the scaling relations \citep{Newman2013}. \\
The BCG centred in the first clump ($\delta R.A., \delta Dec$)= (+20.0", -32.0") can have a core radius between 3.5 and 50.0 kpc, the cut radius can vary from 25 to 320 kpc, the ellipticity from 0.2 to 0.7, the orientation from 120 to 180 degrees and the velocity dispersion from 100 to 500km/s.\\
As for the BCG centred at ($\delta R.A., \delta Dec$)= (-40.0", +40.0"): its core radius can vary between 25 and 35 kpc, the cut radius from 670 to 730 kpc. Its ellipticity is allowed to vary between 0.4 and 0.65 kpc, the orientation angle from 105 to 115 degrees and the velocity dispersion from 80 to 420 km/s.

\subsubsection{NFW }\label{mod3}
This reconstruction, as in \ref{mod1}, has two cluster-scale haloes but modelled using a NFW density profile instead. \\
The clump fixed at ($\delta R.A., \delta Dec$)= (+20.0", -32.0") away from the cluster centre can have an orientation between 130 and 140 degrees, its velocity dispersion can vary between 500.0 and 2000.0 km/s and its scale radius varies from 50 to 280kpc and we let the ellipticity vary from 0.2 to 0.6.\\
The ellipticity of the second clump - fixed at ($\delta R.A., \delta Dec$)= (-40.0", +40.0")- can reach values up to 0.6, the orientation is set between 105 and 115 degrees, the velocity dispersion is set to vary between 500 and 1800km/s and its scale radius between from 50 and 280kpc. \\ 
Even if the ellipticity intervals are quite large for both clumps, after optimization these clumps do no reach values $e\gtrsim0.4$. This is in agreement with \citet{golse2002b} who demonstrated that the NFW profile is ill-defined for large ellipticities as defined in \textsc{Lenstool}. This will remain true for both clusters NFW models.

\subsubsection{NFW - PIEMD} \label{mod4}
We add two BCG components (modelled with a PIEMD profile) to the reconstruction in \ref{mod3}, in the same way as in \ref{mod2}.
At this point, we realized that the NFW models were a better fit to ARES (than PIEMD for the large-scale haloes) as the improvement of the logarithm of the Evidence in Table \ref{table:2} shows (confirmed as the profile used to generate \textit{Ares}'s large-scale haloes upon the unblinding during the challenge). Therefore, the following models, more complex as they also take into account more distant structures or an Hernquist profile, were only modelled using the NFW profile for the large-scale haloes.\\
The last two models are "post-unblinding" models. Once the input convergence map was made available we discovered some structures in the outskirts of the cluster, out of the HST field of view (see Figure \ref{clusters}).

\subsubsection{NFW - PIEMD  + SUBS} \label{mod5}
This model, together with the model \ref{mod7}, includes six additional dark matter components located in the outskirts of the cluster.
This model is exactly like \ref{mod4} with the additional substructures.\\
The substructures are located at ($\delta R.A., \delta Dec$)= (+110.9", -118.4"); (+84.3", -138,8"); (+136.0", -127,98"); (+57.6", -83.2"); (-108,9", +136,9"); (+111.3", -75.0") away from the cluster's centre. They are modelled with NFW profiles, considered as spherical, their velocity dispersions are allowed to vary between 100 and 500km/s and their scale radius between 20 and 200 kpc.  \\

\subsubsection{NFW - HERNQUIST} \label{mod6}
The large-scale clumps are modelled as in \ref{mod4}. 
The BCGs, on the other hand have a different density profile: the Hernquist profile as described in Section \ref{massmod}. \\
The BCG located at ($\delta R.A., \delta Dec$)= (+20.0", -32.0") has a core radius between 1.0 and 40.0 kpc. Its ellipticity can reach values up to 0.4, the orientation varies between 90 and 180 degrees and the velocity dispersion can vary from 100 to 400 km/s. \\
For the BCG located at ($\delta R.A., \delta Dec$)= (-40.0", +40.0") the core radius is allowed to vary between 1.0 and 40.0 kpc, the ellipticity from 0.1 to 0.7, the orientation from 90 to 180 degrees and the velocity dispersion between 90 and 400 km/s. 

\subsubsection{NFW - HERNQUIST +SUBS} \label{mod7}
For this model, the large-scale haloes and BCGs components are modelled as in \ref{mod6} and the distant substructures as in \ref{mod5}.

\subsubsection{NFW - PIEMD + shapes} \label{mod8}
Last, we consider how taking into account the shapes of the galaxy-scale haloes impacts the reconstruction. The large-scale haloes and BCGs components are modelled as in \ref{mod4} but we use SExtractor \citep{Bertin1996} to measure the semi-major and semi-minor axis of the fitted ellipsis of each cluster galaxy which are then taken into account in the modelling. 

\subsection{\textit{Hera}} \label{[section5]}
In spite of the unimodal appearance of the cluster and its convergence contours in the bottom panel of Figure \ref{clusters}, Hera is also modelled as a bimodal cluster. Indeed, if this cluster is fitted with only one dark matter central clump the $\chi^{2} $ is twice larger.\\
The cluster members of the input catalogue were taken into account with $\mathrm{m_{F814W}}$ < 24.0 and were modelled in the same way for all models: parametrized by a PIEMD profile with a fixed core radius of 0.15 kpc, a velocity dispersion $\sigma^{\star}$ allowed to vary between 60.0 and 100.0 km/s, a cut radius $r_{cut}^{\star}$ with values from 1.0 kpc to 200.00 kpc, considered spherical, with $mag0=19.8$ and following the scaling relations \citep{Faber1976}. \\
Moreover, in the same way as for \textit{Ares}, four massive galaxies were more carefully modelled (indicated as green diamonds in Figure \ref{clusters}), using a PIEMD density profile and fitted in the same way for each model.\\ 
The galaxy labelled as 1 in Figure \ref{clusters}  is located at ($\delta R.A., \delta Dec$)= (+25.91", +26.98") away from the cluster centre,considered spherical, an angle position of 72 degrees and a core radius of 0.07 kpc. Its cut radius $r_{cut}$ can vary from 10.0 to 100.0 kpc and its velocity dispersion $\sigma$ from 50.0 to 200.0 km/s.\\
The next one, labelled as 2 in Figure \ref{clusters}, is positioned at ($\delta R.A., \delta Dec$)= (+60.38", +20.2"),  with a core radius of 0.31 kpc. Its ellipticity can go up to 0.5, the angle position is set between -65.0 and 90.0 degrees, the cut radius $r_{cut}$ can vary from 30.0 to 400.0 kpc and its velocity dispersion $\sigma$ from 50.0 to 500.0 km/s.\\
The third halo of a massive galaxy (labelled as 3 in Figure \ref{clusters}) is located at ($\delta R.A., \delta Dec$)= (-19.665", -11.078") with a core radius of 0.1 kpc. Its ellipticity can vary between 0.1 to 0.7, the angle position from 22 to 90 degrees, the cut radius $r_{cut}$ is allowed to vary between 8.0 to 150 kpc and the velocity dispersion $\sigma$ from 400 to 200.0 km/s. \\
Finally, the last one is a very massive galactic halo (labelled as 4 in Figure \ref{clusters}) having a significant impact in the total mass distribution profile. This halo is located at ($\delta R.A., \delta Dec$)= (-7.0", -46.0") with a core radius of 0.1, an ellipticity set between 0.1 and 0.7, an angle position allowed to vary between 75.0 to 90.0 degrees, a cut radius $r_{cut}$ between 8.0 and 250.0 kpc and a velocity dispersion $\sigma$ between 40.0 and 600.0 km/s.\\
All the multiple images provided were included in the model with a positional uncertainty of 0.5" and the optimization is performed in the source plane. As for \textit{Ares} we checked that image plane and source plane models were giving similar results.

\subsubsection{PIEMD  }\label{mod1h}
This model contains two large-scale halos whose coordinates are placed at ($\delta R.A., \delta Dec$)= (0.00", 0.00") and ($\delta R.A., \delta Dec$)= (+20.0", +2.00") away from the centre of the cluster respectively (and no BCGs). They are allowed to move around by $\pm$ 2.00". \\
For the first clump, we let the core radius vary from 5 kpc to 37 kpc with a fixed cut radius of 3000 kpc. The ellipticity of this halo is allowed to vary values from 0.0 and 0.7, its orientation angle from 0.0 to 180.0 degrees and its velocity dispersion can vary from 400 to 1500 km/s.
The second dark matter clump can have a core radius with any value from 5 kpc to 40 kpc and cut radius, also fixed, of 3000 kpc. Its ellipticity can take any value from 0.0 to 0.7, its orientation can vary from 0.0 to 180 degrees and the velocity dispersion can vary from 600 to 1000 km/s.

\subsubsection{PIEMD - PIEMD }\label{mod2h}
We add two BCG components (modelled with a PIEMD profile) to the reconstruction in \ref{mod1h}. \\
The BCG fixed at ($\delta R.A., \delta Dec$)= (0.0", 0.0") has a fixed core radius of  0.5 kpc and a cut radius that can vary from 30.0 to 450.0 kpc. Its ellipticity can go from 0.0 to 0.4, the orientation from 5 to 90 degrees and the velocity dispersion from 100 to 500 km/s. \\
The ($\delta R.A., \delta Dec$)= (+20.0", +2.00") centred BCG has a fixed core radius of 1.0 kpc, the cut radius can vary from 20.0 to 400.0 kpc, the ellipticity from 0.0 to 0.3, the orientation angle from 12.0 to 90.0 degrees and the velocity dispersion between 100 and 500 km/s. 

\subsubsection{PIEMD - HERNQUIST }\label{mod3h}
In this reconstruction, the two added BCGs to the model \ref{mod1h} are modelled using an Hernquist density profile. \\
The BCG located at ($\delta R.A., \delta Dec$)= (0.0", 0.0") has a fixed core radius of 0.5kpc. Its ellipticity can reach values up to 0.4, the orientation angle is allowed to vary from 10 to 90 degrees and the velocity dispersion from 100 to 500 km/s. \\
For the BCG located at ($\delta R.A., \delta Dec$)= (+20.0", +2.00"), the core radius is fixed to 1.0 kpc, the ellipticity can vary from 0.0 to 0.3, the orientation angle from 6 to 90 degrees and the velocity dispersion between 100 and 500 km/s. 

\subsubsection{NFW  }\label{mod4h}
This model, as in \ref{mod1h}, has two large-scale haloes but modelled using a NFW density profile instead. \\
For the clump located at ($\delta R.A., \delta Dec$)= (0.0", 0.0"), aligned with the centre of the cluster we let the ellipticity vary from 0.0 to 0.7, its orientation from 50 to 180 degrees. It can have a velocity dispersion between 500 and 1500 km/s and scale radius varying from 50 to 280 kpc.\\
The ellipticity of the second clump (located at ($\delta R.A., \delta Dec$)= (+20.0", +2.00") can reach a value from 0.1 up to 0.7, the orientation is set between 0.0 and 180 degrees, the concentration between 50 and 1500 km/s and its scale radius from 50 to 280 kpc.

\subsubsection{NFW - PIEMD} 
For this model, the large-scale haloes are modelled as in \ref{mod4h} and BCGs components are modelled as in \ref{mod2h} 

\subsubsection{NFW - HERNQUIST}
For this last model, the large-scale haloes are modelled as in \ref{mod4h} and BCGs components are modelled as in \ref{mod3h} 
\section{Estimators} \label{[results]}
In this section we present four estimators used to assess the quality of our reconstructions and quantify the impact of the systematic errors arising from the choice of density profiles and configurations in the modelling of both \textit{Ares} and \textit{Hera}.

	\begin{figure*}
	\centering
	\includegraphics[width=\linewidth]{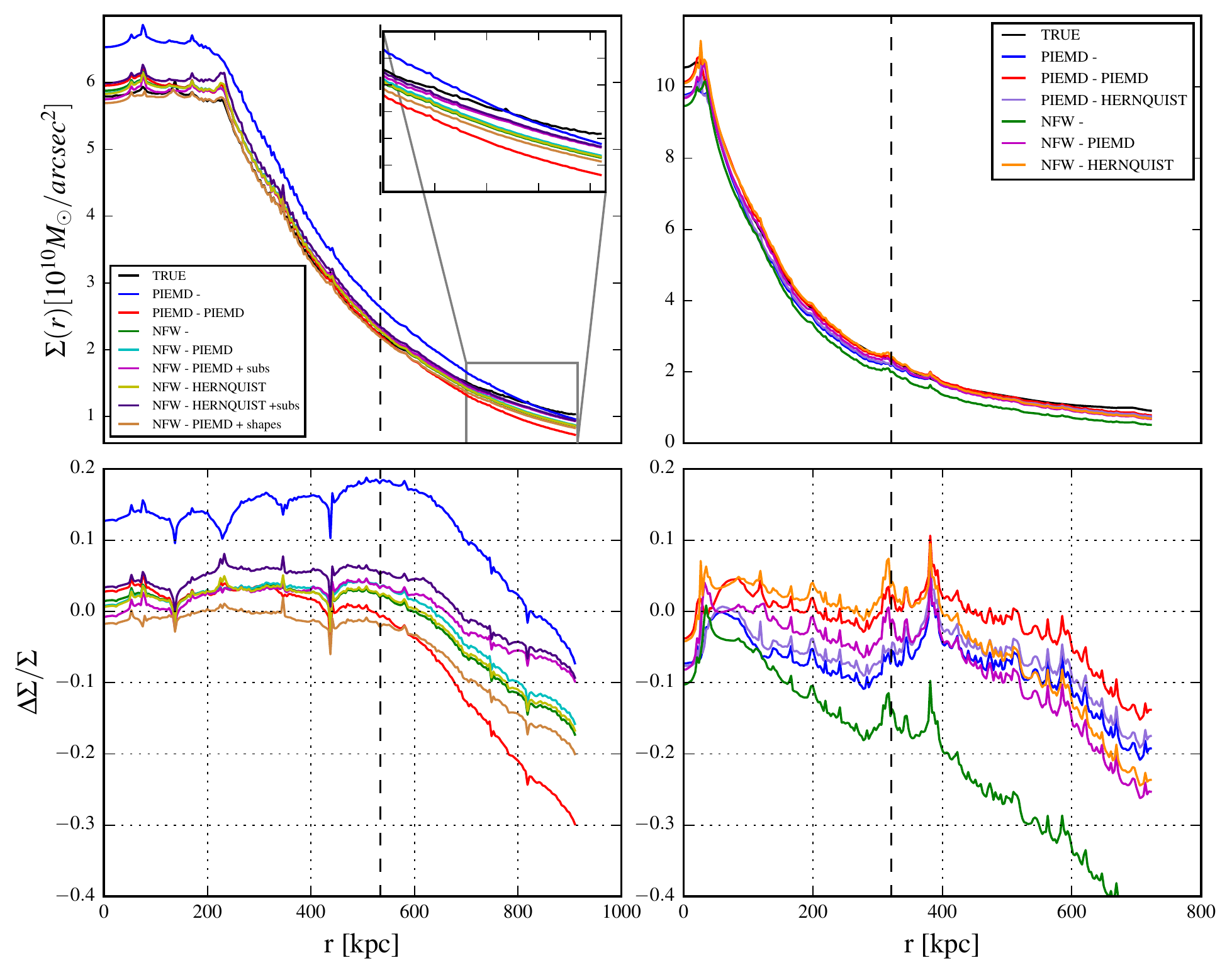} 
	\caption{Top panels: Density profiles for each model of Section \ref{[section4]} for \textit{Ares} (left) and of Section \ref{[section5]} for Hera (right). Bottom panels: Relative error of each model on the density for \textit{Ares} (left) and \textit{Hera} (right). The vertical dashed lines represent the radius below which there are multiple images.\\}
	\label{dens_ares}
    \end{figure*} 

	\begin{figure}
	\centering
	\includegraphics[width=\columnwidth]{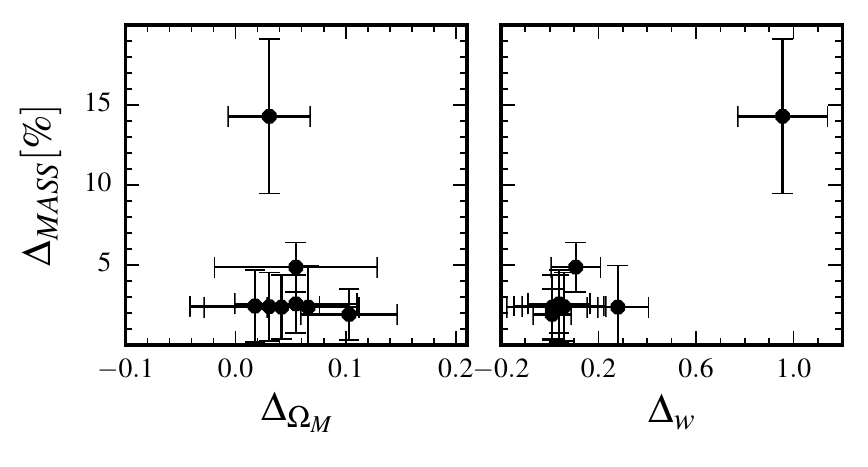} 
	\caption{Bias on the total mass inside the radius below which there are multiple images for each model of Section \ref{[section4]} for \textit{Ares} as a function of the bias on the estimations of the cosmological parameters $\mathrm{\Omega_{M}}$ (left panel) and $w$ (right panel).}
	\label{biasmass}
    \end{figure}

	\begin{figure*}
	\centering
	\includegraphics[width=0.8\linewidth]{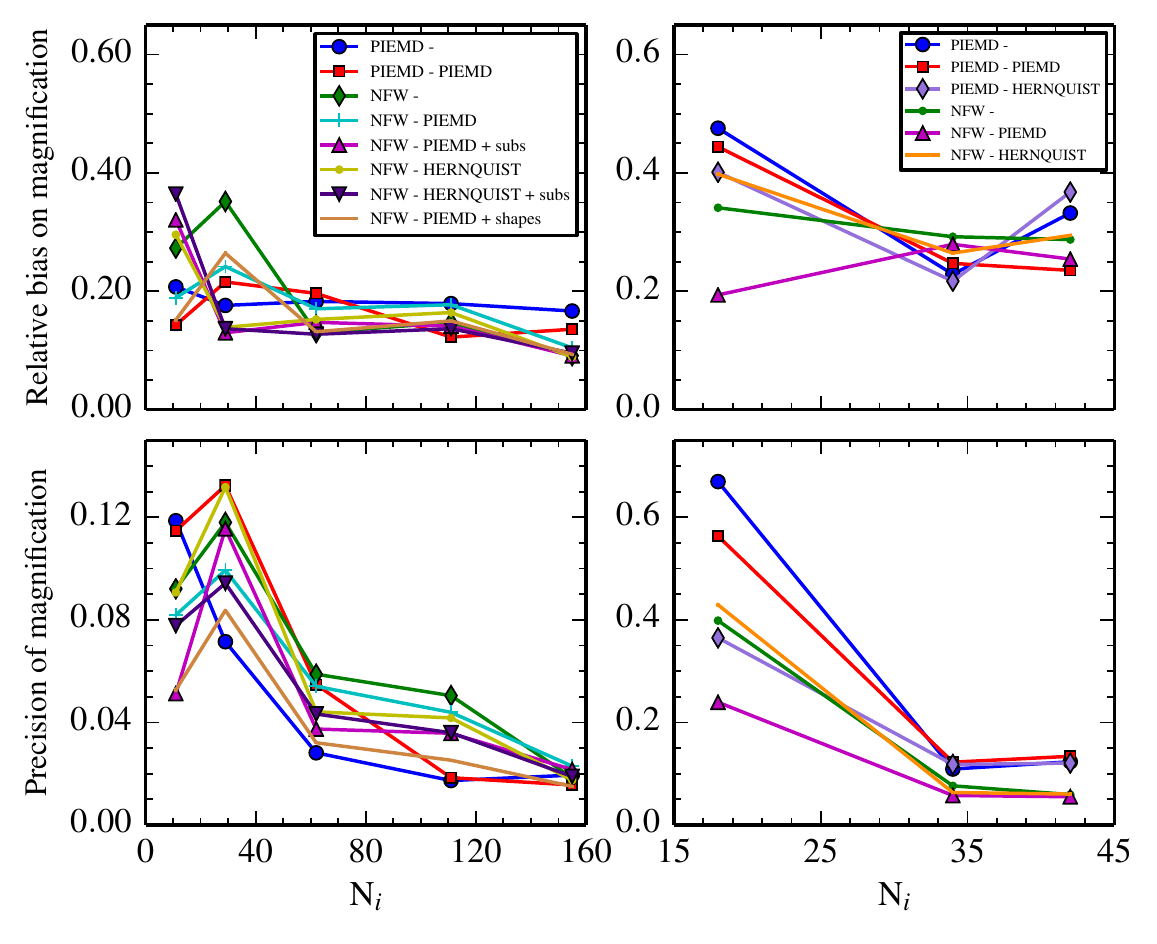} 
	\caption{Top panels: Relative error on magnification as a function of the increasing number of multiple images taken into account in the modelling. Left: \textit{Ares}. Right: \textit{Hera}.\\
	Bottom panels: Relative precision on the magnification measurements as a function of the increasing number of multiple images. Left: \textit{Ares}. Right: \textit{Hera}.}
	\label{magni}
\end{figure*}

\subsection{Statistical quality assessment}\label{stat}

We use several statistical estimators to assess the quality of each reconstruction: the logarithm of the Likelihood, the RMS (in arcseconds) in the image plane, the reduced $\chi^{2}$, the logarithm of the Evidence and the Bayesian Information Criterion \citep[following][]{Lagattuta2016}.
The results are summarized in Table \ref{table:2} and Table \ref{table:3} for Ares and Hera respectively. \\ 
We would like to draw the reader's attention to the fact that all estimators are consistent with each other but using the logarithm of the Evidence appears to be a better way to differentiate between models. On the other hand, the RMS does not necessarily reflect the improvement of the reconstruction. However, both the BIC and AICc strongly penalize models including the distant substructures in the modelling. Indeed, the greater number of free parameters in these models would not be justified be even though they improve the mass distribution at large radii as well as tighten the cosmological constraints (see Figures \ref{dens_ares} and \ref{cosmotousares}, respectively). \\
As expected, in the case of \textit{Ares}, considering a NFW profile for the large-scale potentials improves the modelling and Model 7 (NFW - HERNQUIST + subs) gives the best fit for this cluster as it is the closest to the true mass distribution. We show that taking into account the substructures in the cluster's outskirts for \textit{Ares} leads to an improvement in the modelling of $\sim20\%$. Wide field imaging around strong lenses is thus crucial in order to detect such structures and include them in future reconstructions.\\
For the model \textit{NFW-PIEMD + shapes}, we see that this additional information worsens the modelling of \textit{Ares} with a lower value of $log(Evidence)$ than the model without. This is in agreement with the true model as input cluster galaxies are spherical \citep{meneghetti2016} thus showing that our modelling technique is sensitive to the shapes of cluster members.
As for \textit{Hera}, the resulting fit is similar for all models. As expected, the reduced $\chi^{2}$ values are larger than those for \textit{Ares}, as the latter was modelled parametrically (assuming that light traces mass) thus better suited for our modelling technique.
%-------------------------------------------------------------------
%\begin{landscape}
	\begin{table*}
	\caption{Mean value of the logarithm of the likelihood and image plane RMS("), reduced  $\chi^{2}$, the logarithm of the Evidence and the BIC (Bayesian information criterion) for different \textit{Ares} strong lensing models. In the "Model" column we specify the density profile chosen for the dark matter haloes and BCGs respectively. Finally we also show the best-fit values of the cosmological parameters $\mathrm{\Omega_{M}}$ and $w$. More details of these models are given in Section \ref{[section4]}.}            
	\label{table:2}      % is used to refer this table in the text
	\centering                          % used for centering table
	\begin{tabular}{c c c c c c c c c}        % centered columns (4 columns)
		\hline\hline      
		\\           % inserts double horizontal lines
		 Model & log(Likelihood) & RMS(") & Reduced $\chi^{2}$ & log(Evidence) & BIC & AICc & $\mathrm{\Omega_{M}}$ & $w$ \\    % table heading 
		\\
		\hline          % inserts single horizontal line
		\\  PIEMD - no BCGs  & 76.61 & 1.80 & 1.27 & 33.51 & -15.23& -24.44& $0.24^{+0.04}_{-0.03}$&$-1.57^{+0.15}_{-0.23}$\\
		\vspace{0.01cm}\\      % inserting body of the table
		 PIEMD - PIEMD & 117.00 & 0.66 & 0.81 & 61.46 & -32.77& -37.93 &$0.28^{+0.10}_{-0.06}$&$-1.02^{+0.05}_{-0.02}$\\
		\vspace{0.01cm}\\
		 NFW - no BCGs  & 144.83 &  0.60& 0.61 & 95.78& -163.17&-97.35 &$0.31^{+0.06}_{-0.06}$&$-1.09^{+0.10}_{-0.24}$\\
		\vspace{0.01cm}\\
		 NFW - PIEMD & 155.28 & 0.73 & 0.58 & 100.70& -109.33 &-83.76&$0.24^{+0.05}_{-0.05}$&$-1.03^{+0.06}_{-0.14}$\\
		\vspace{0.01cm}\\
	    NFW - PIEMD + SUBS & 166.53 & 0.55 & 0.48 & 107.65& -45.59&-47.14&$0.19^{+0.06}_{-0.03}$&$-0.99^{+0.05}_{-0.10}$\\
		\vspace{0.01cm}\\
		 NFW - HERNQUIST & 145.54 & 0.58& 0.61 & 93.58&-118.59 &-78.97& $0.27^{+0.03}_{-0.11}$&$-1.05^{+0.06}_{-0.27}$\\
		\vspace{0.01cm}\\
		 NFW - HERNQUIST + SUBS & 168.31  & 0.64 &0.50 &111.43 &-60.65 &-55.69& $0.33^{+0.03}_{-0.04}$&$-1.25^{+0.13}_{-0.12}$\\
		\vspace{0.01cm}\\
		 NFW - PIEMD + shapes & 135.56 & 0.57 & 0.65 & 79.82& -87.14& -64.04&$0.22^{+0.10}_{-0.05}$ & $-0.90^{+0.07}_{-0.12}$
		\\ 
		\hline\hline                                 %inserts single line
	\end{tabular}
\end{table*}
%\end{landscape}
\begin{table*}
	\caption{Same as Table \ref{table:2} for the \textit{Hera} cluster. More details of these models are given in Section \ref{[section5]}.}            
	\label{table:3}      
	\centering                         
	\begin{tabular}{c c c c c c c c c}        
		\hline\hline     
		\\            
		 Model & log(Likelihood) & RMS(") & Reduced $\chi^{2}$ & log(Evidence) & BIC &AICc & $\mathrm{\Omega_{M}}$ & $w$ \\   
		\\ 
		\hline         
		\\  PIEMD - no BCGs &  -35.93 & 0.99  &  2.85 & -77.20 & 207.51&126.42&$0.54^{+0.14}_{-0.20}$&
		$-1.15^{+0.18}_{-0.73}$\\      
		\vspace{0.01cm}\\
		 PIEMD - PIEMD& -19.44 & 1.21 & 2.64 & -64.97 & 210.71 &151.36& $0.41^{+0.12}_{-0.09}$&$^-1.55^{+0.58}_{-0.09}$\\
		\vspace{0.01cm}\\
		 PIEMD - HERNQUIST & -31.80 & 0.96 & 2.98 & -71.60& 226.38& 152.24&$0.44^{+0.16}_{-0.20}$	&$-1.05^{+0.15}_{-0.57}$\\
		\vspace{0.01cm}\\
		 NFW - no BCGs  &-43.44 & 0.98&2.71 &-79.81 & 222.53&133.93 &$0.71^{+0.09}_{-0.30}$&$-1.30^{+0.64}_{-0.37}$\\
		\vspace{0.01cm}\\
		 NFW - PIEMD & -46.47 & 1.00 & 3.03 & -85.40 & 264.77& 178.39 &$0.33^{+0.18}_{-0.10}$ &$-1.19^{+0.21}_{-0.50}$\\
		\vspace{0.01cm}\\
		 NFW - HERNQUIST & -44.59 & 0.99& 2.96 & -81.40& 251.96 & 165.02 &$0.32^{+0.22}_{-0.07}$&$-1.48^{+0.47}_{-0.24}$\\
		\hline\hline                   
	\end{tabular}
\end{table*}
%-------------------------------------------------------------------
%-------------------------------------------------------------------
\subsection{Density profiles}\label{density}
In this section we present the comparison between the radial density profile for all of our models and the true one.\\
In order to compare our projected mass maps to the input convergence map (with sources assumed at $z_{s}$=9), the latter is normalised by multiplying the value of every pixel by $\Sigma_{crit}$ (at $z_{s}=9$, and assuming the input cosmology) in order to have the associated surface mass density $\Sigma(x,y)$ as:
\begin{equation}
\centering
{\Sigma(\vec{\theta})} = \kappa(\vec{\theta}) {\Sigma_{crit}}\mathrm{,}
\end{equation}
where:
\begin{equation}
\centering
\Sigma_{crit} = \dfrac{c^{2}}{4\pi G}\dfrac{D_{s}}{D_{l}D_{ls}}\mathrm{.}
\end{equation}
with $c$, the speed of light, $G$ the  Newtonian constant and $D_{s}, D_{l}$ and $D_{ls}$ the  angular  diameter  distances  between observer-source,  observer-lens  and  source-lens, respectively.\\
Our maps and the true one have the same field of view and spatial resolution. \\
For each model of Section \ref{[SLM]}, we compute the radial density profile and compare it to the true profile in Figure \ref{dens_ares}.\\
\begin{enumerate}
	\item \textit{Ares}:\\
	 Regardless of the profile used for the large-scale potentials and BCGs, the mass distribution is well constrained within $\sim5\%$ inside the radius below which there are constraints-defined as the radius of the circle enclosing all multiple images (except for the model PIEMD, which is overestimating the cluster's density by $\sim14\%$). However, the density profile in the outskirts of the \textit{Ares} cluster (beyond the black vertical dashed line, representing the radius containing multiple images) tends to be underestimated by up to $\sim30\%$. Including the substructures in the modelling helps to better constrain the mass distribution, with an improvement between $\sim5\%$ to $\sim20\%$ in the most distant regions of the cluster. On average, the statistical error is about $5\%$ of the signal for all models, although for Model 1 (PIEMD), it varies from $4\%$ in the inner regions to up to $8\%$ in the outskirts.\\
	 We also display in Figure \ref{biasmass} the bias on the estimation of $\mathrm{\Omega_{M}}$ and $w$ (i.e the absolute error between the true value and the mode of the distribution for each of our models, see Figure \ref{cosmotousares}) depending on the bias on the total mass (inside the radius below which there are multiple images) for all of our models of Section \ref{[section4]} for \textit{Ares} (as the constraints on $\mathrm{\Omega_{M}} - w$ for \textit{Hera} in Figure \ref{cosmotoushera} are too wide to draw any conclusion). As expected the larger the bias on the mass is, the larger the bias on the cosmological parameters. The bias on the mass is therefore a cosmological quality estimator. Also we show that the equation of state parameter $w$ is more affected by a larger bias on the cluster mass than $\mathrm{\Omega_{M}}$. This result is in agreement with \citet{golse2002} where the authors showed that the matter density parameter $\mathrm{\Omega_{M}}$ can be better constrained than $\Omega_{\Lambda}$ for a set of simple mock strong lenses. In this work we confirm that $\mathrm{\Omega_{M}}$ is less sensitive to the modelling than $w$, even for more complex clusters such as \textit{Ares}.
	\item \textit{Hera}:\\ As expected, Hera is less constrained inside the radius below which there are multiple images (within $\sim5-15\%$) and the density profile at the very core of the cluster differs significantly from the true distribution. The density profile is well constrained by all models up to the radius containing multiple images. In the outskirts of the cluster, the density profile differs from the true distribution from $\sim10\%$ up to $\sim40\%$ depending on the model. The statistical error is larger than for Ares, especially for models 3, 4 and 6 , for which it varies from  $\sim5\%$ in the center up to $10-19\%$ in the outskirts.
\end{enumerate}

\subsection{Relative bias on magnification}
In this section, we detail how we compute the magnifications from our lens models and compare them to the true values. \\
If we consider sources smaller than the angular scale on which the lens properties change, the Jacobian matrix describing the distortion (in shape and size) of images is then:\\
\begin{equation} \label{eq_mag1}
\mathcal{A}(\theta)= \begin{pmatrix}
\centering
1 - \kappa - \gamma_{1} & \gamma_{2} \\ 
\gamma_{2} & 1 - \kappa + \gamma_{1} \\
\end{pmatrix}\mathrm{,}
\end{equation} 
where $\kappa$ is the convergence and $\gamma_{1}$, $\gamma_{2}$ the first and second shear components respectively.
The magnification is then the inverse of the determinant of $\mathcal{A}$:
\begin{equation}\label{eq_mag2}
\centering
\mu = (det\mathcal{A})^{-1} = \dfrac{1}{| (1 - \kappa)^{2} - |\gamma|^{2} |}\mathrm{,}
\end{equation}
where $mu$ is the magnification of the source.
\begin{enumerate}
	\item \textit{True magnification}\\
To obtain the true magnification for each multiple image we use the true convergence and shear maps at  $z_{s} = 9$ at any location in the image plane covering a field of view of $300"\times 300"$ for \textit{Ares} and $227.16" \times 227.16"$ for \textit{Hera}.\\
However, to have the true magnification at the source's redshift, we multiply the true convergence $\kappa$ and shear components $\gamma_{1}$, $\gamma_{2}$ by a normalizing factor:\\
\begin{equation}\label{eq_mag3}
\centering
\dfrac{D_{OS}}{D_{LS}}(z_{s}=9) \times \dfrac{D_{LS}}{D_{OS}}(z_{s}=z)\mathrm{.}
\end{equation} 
Finally, we interpolate this map to get the magnification at the position of all input multiple images. \\ 
	\item \textit{Measured magnification}\\
The magnification for each arclet is measured for all of our reconstructions with the \textsc{Lenstool} software using equations \ref{eq_mag1} and \ref{eq_mag2}. 
\end{enumerate}
We are interested in how the measured magnification evolves with an increasing number of multiple images.
To do so, we run SExtractor \citep{Bertin1996} to extract and measure the magnitudes on the F814W image (deeper thus more arcs detected) of all the multiple images which are then divided into five different catalogues with increasing magnitude thresholds in the F814W filter (being uniformly distributed in the field). \\
We then run each lens model presented in Sections \ref{[section4]} and \ref{[section5]} with all of the new multiple image catalogues (with different magnitude thresholds). We compute the relative bias on magnification per multiple image as well as the 25-th and 75-th percentiles of this distribution. Finally for each magnitude catalogue we compute an average bias on the magnification with:\\
\begin{equation}
\delta \mu= p75\left(\dfrac{\mu^{fit}-\mu^{true}}{\mu^{true}}\right) - p25\left(\dfrac{\mu^{fit}-\mu^{true}}{\mu^{true}}\right)\mathrm{,}
\end{equation}
where $\mu^{fit}$ is the mode of the magnification distribution per catalogue.\\
We show in Figure \ref{magni} the relative bias on magnification as a function of the number of images taken into account in the modelling for both clusters (top panel) and the precision of our measurements (bottom panel). The bias on the magnification is of the order of $\sim20\%$ for \textit{Ares}. This bias can go up to $40\%$ for the smallest bins. The magnifications of the multiple images in Hera are less constrained with an average bias of $\sim30\%$. The relative bias on magnification is not reduced by an increasing number of constraints for either of the clusters. However, the precision on magnification improves by a factor $\sim 4$ when increasing the constraints in the modelling as stated in \citet{Jauzac2014}. All models provide similar measurements of the magnification and modelling galaxy clusters in a more complex way is not translated into an improvement in the accuracy or precision of magnification. \\
Taking into account the substructures in the cluster's outskirts leads to a very localised improvement of the magnification bias of $\sim15\%$ but remains constant within the cluster's core.\\
%In \citet{Johnson2016} the authors analyse the systematic uncertainties arising from the selection of constraints in the \textit{Ares} cluster. They find that, by comparing models having different constraint selections with their fiducial model (all multiple images with spectroscopic redshifts), there is a clear trend of decreasing systematic errors with an increasing number of multiple images but remain constant at $N\ge25$.

\subsection{Cluster ellipticity and orientation}
\begin{figure*}
		\centering
		\includegraphics[width=\linewidth]{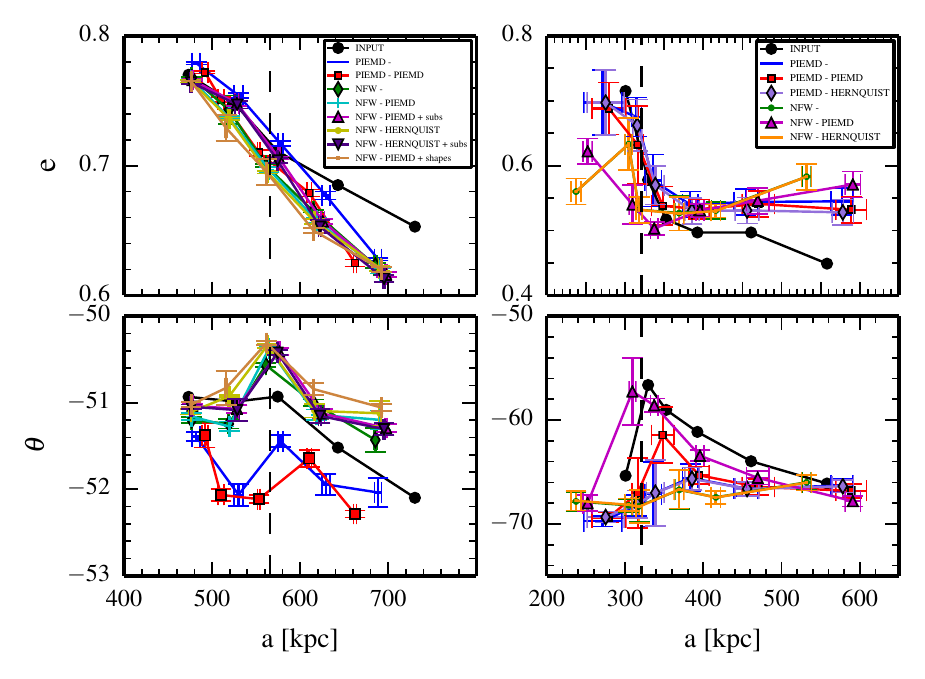} 
		\caption{Top panels: radial profile of the cluster's ellipticity. Bottom panels: radial profile of the orientation angle. The dashed vertical lines represents the limit for which we have multiple images. Left panels: \textit{Ares} cluster. Right panels: \textit{Hera} cluster.}
		\label{ellip}
	\end{figure*}

We use the mass maps generated for each model and the normalised true convergence map (as explained in \ref{density}) to compute the cluster ellipticity as a function of the major semi-axis and its position angle for each model of Section \ref{[SLM]}.\\ 
We set a list of iso-surface density thresholds to fit an ellipse to each contour. As both clusters are bimodal, the highest iso-density contour is set to be the one that encloses both clumps. \\
The ellipticity is then computed as follows:
\begin{equation}
\centering
e = \dfrac{a^{2} - b^{2}}{a^{2} + b^{2}}\mathrm{,}
\end{equation}
where a and b are the measured major and minor semi-axes respectively.
The radial profiles of the ellipticity and of the orientation angle for both clusters are shown in Figure \ref{ellip}.\\
For each panel the true values are given by the black line. Overall, the values of all the models are in good agreement with the true value in the inner regions (before the dashed vertical line, representing the radius up to where we have multiple images) but tend to differ in the outskirts as our models do not have constraints anymore, thus extrapolating from the core region. The ellipticity in the outskirts of \textit{Ares} is underestimated but overestimated for \textit{Hera}. The values of the cluster's orientation angle are recovered within 5 degrees for \textit{Ares} and within 10 degrees for \textit{Hera}.\\
This estimator does not allow to discriminate between the different models as all models are in perfect agreement.
	
\section{Cosmography}\label{cosmosec}
In this section we investigate the impact of considering different cluster modellings on the estimation of cosmological parameters. \\
We assess first if the choice of different density profiles for the cluster components (large-scale potentials and BCGs) has a significant impact on constraining the $\mathrm{\Omega_{M}}$ and $w$ parameter space.\\
We also analyse the systematic error introduced when using different redshift catalogues for the background sources (spectroscopic only, spectroscopy and photometry) for the lens modelling. 

\subsection{Estimation of $\mathrm{\Omega_{M}}$ and $w$ } \label{cosmo1}
\begin{figure*}
	\includegraphics[trim=5cm 2cm 2.5cm 4cm, width=4cm]{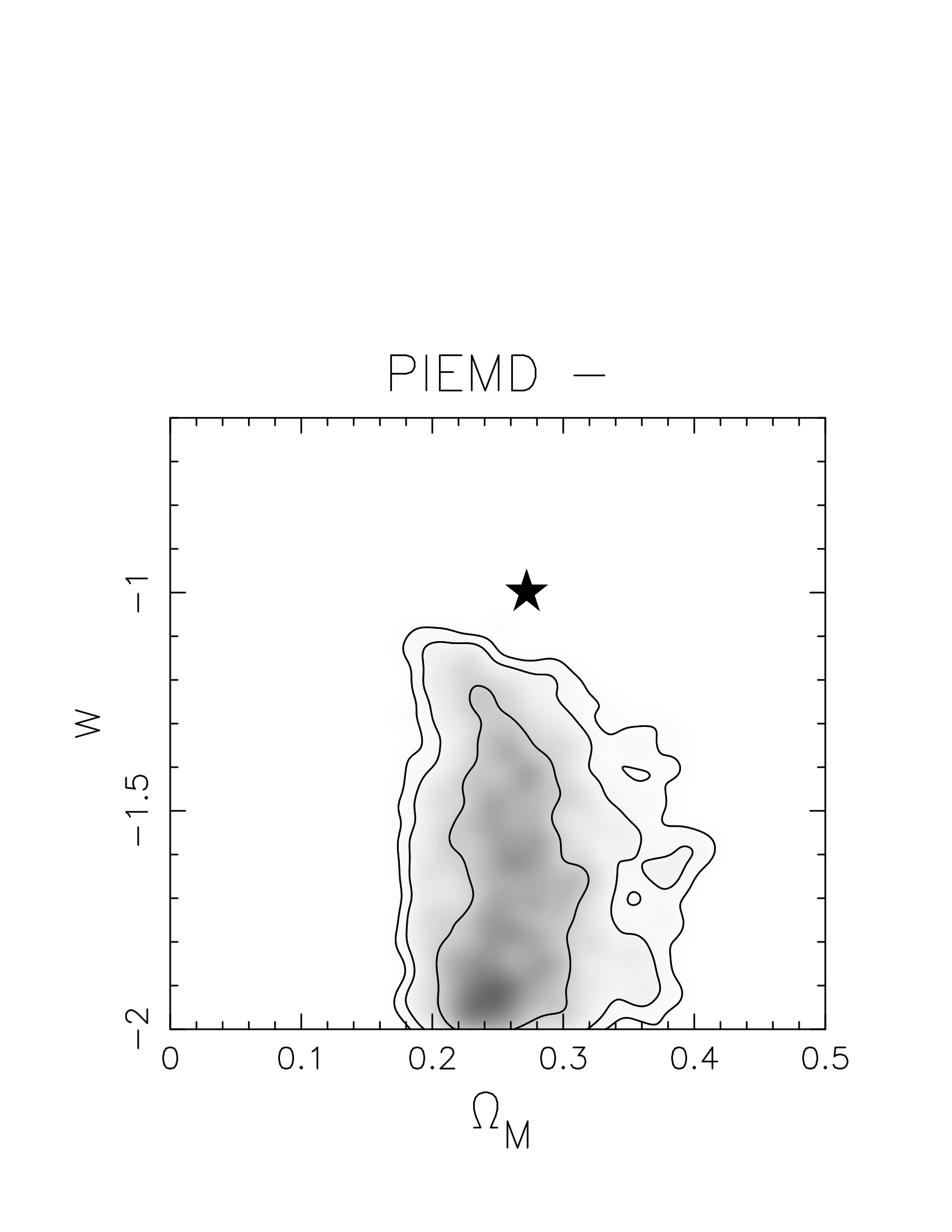} 
	\includegraphics[trim=5cm 2cm 2.5cm 5cm, width=4cm]{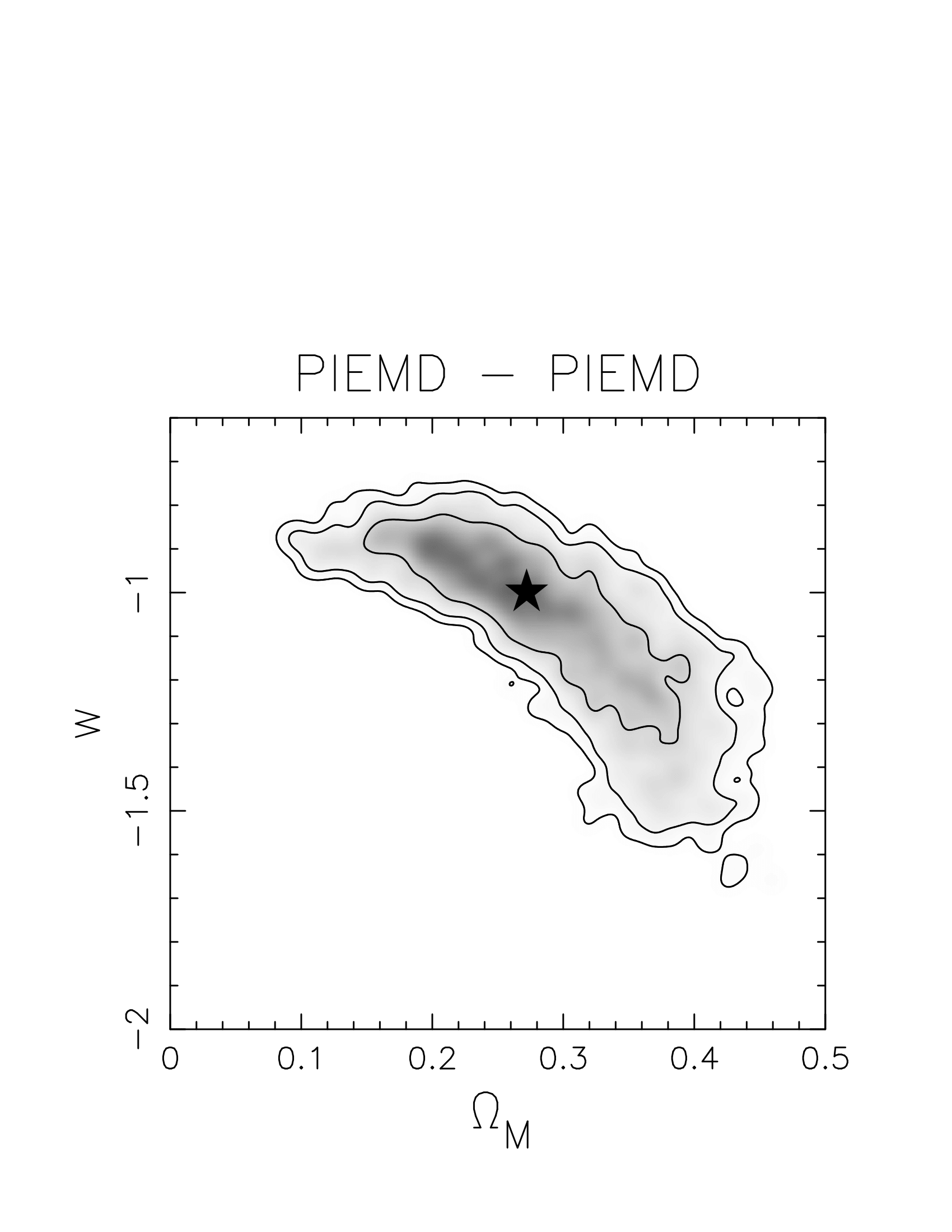}
	\includegraphics[trim=5cm 2cm 2.5cm 5cm, width=4cm]{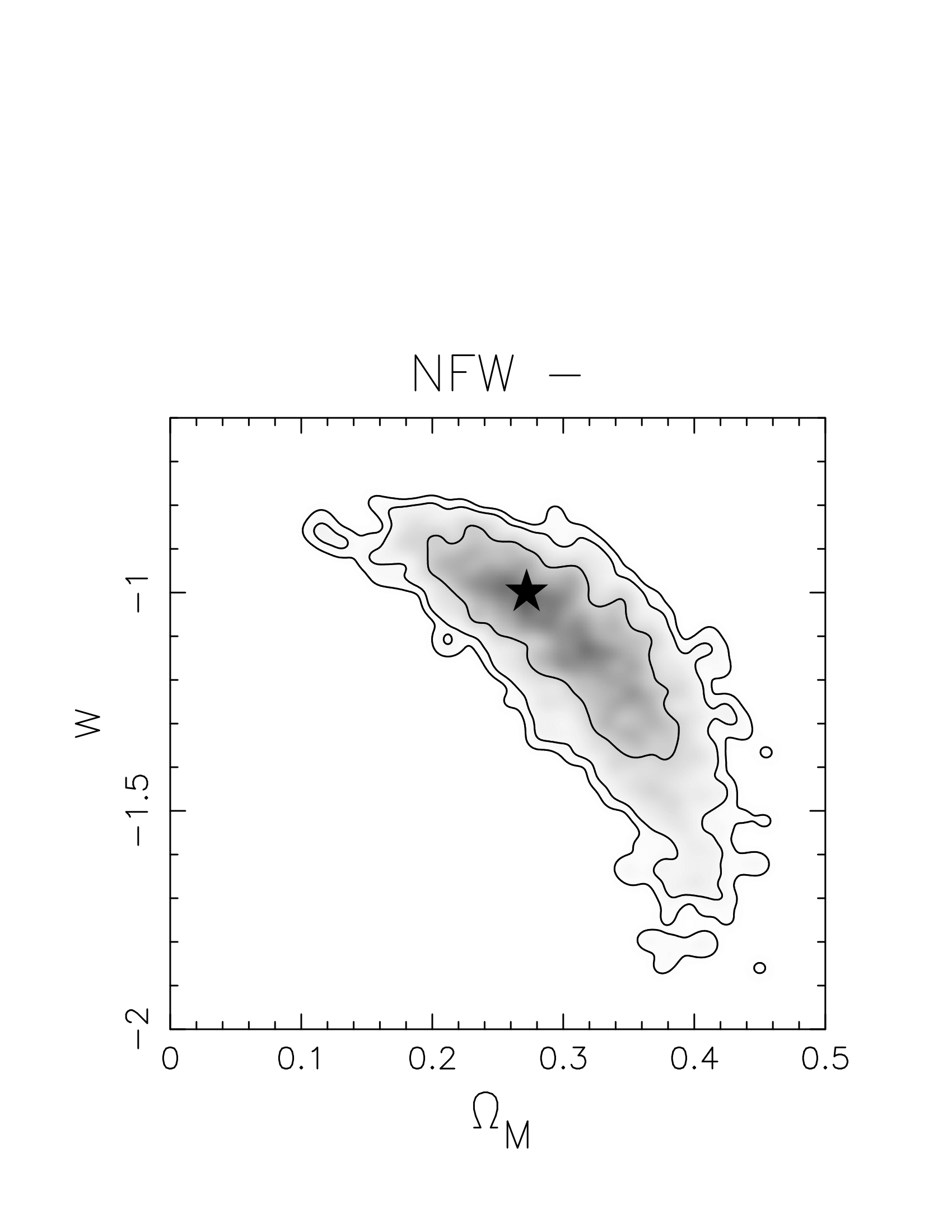}
	\includegraphics[trim=5cm 2cm 2.5cm 5cm, width=4cm]{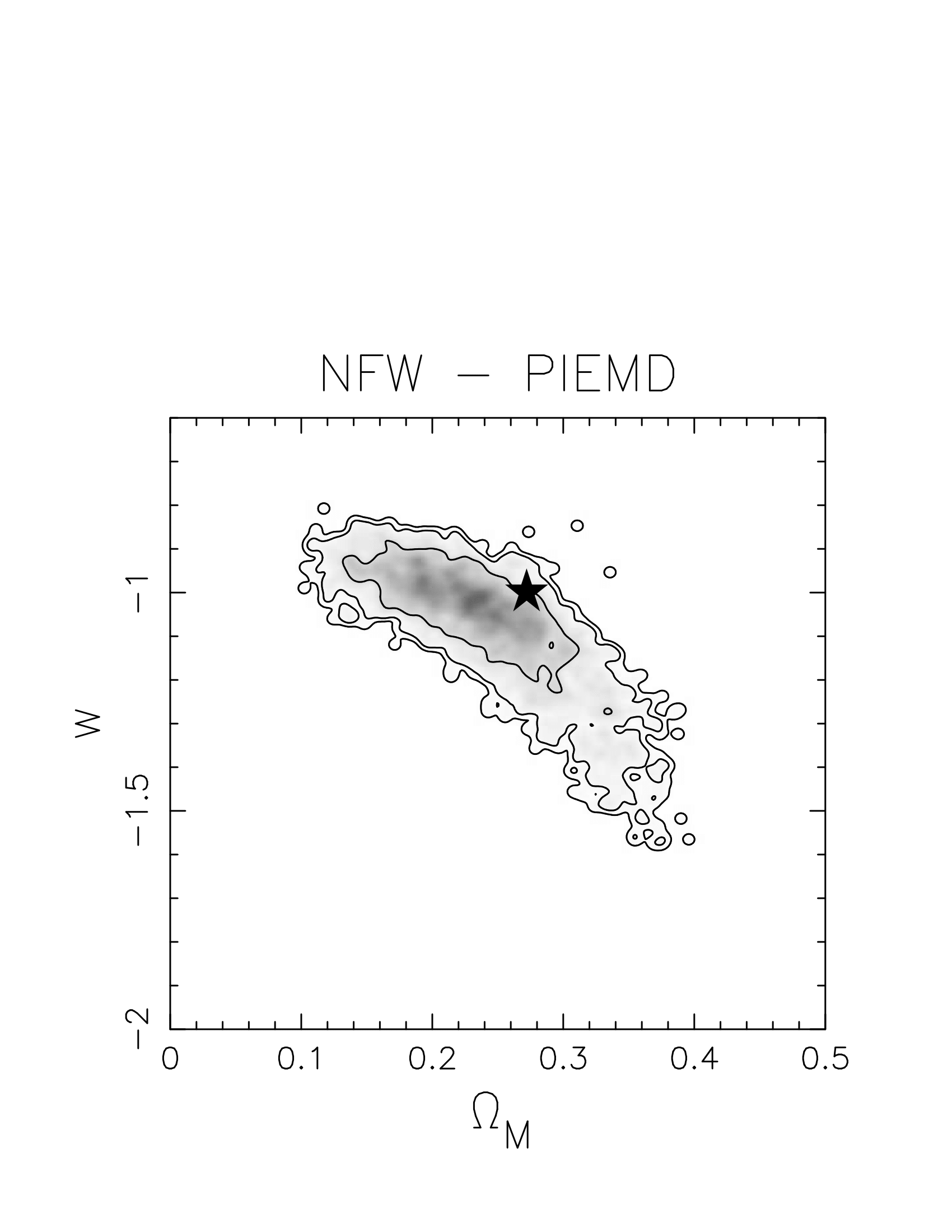}
	\includegraphics[trim=5cm 2cm 2.5cm 4cm, width=4cm]{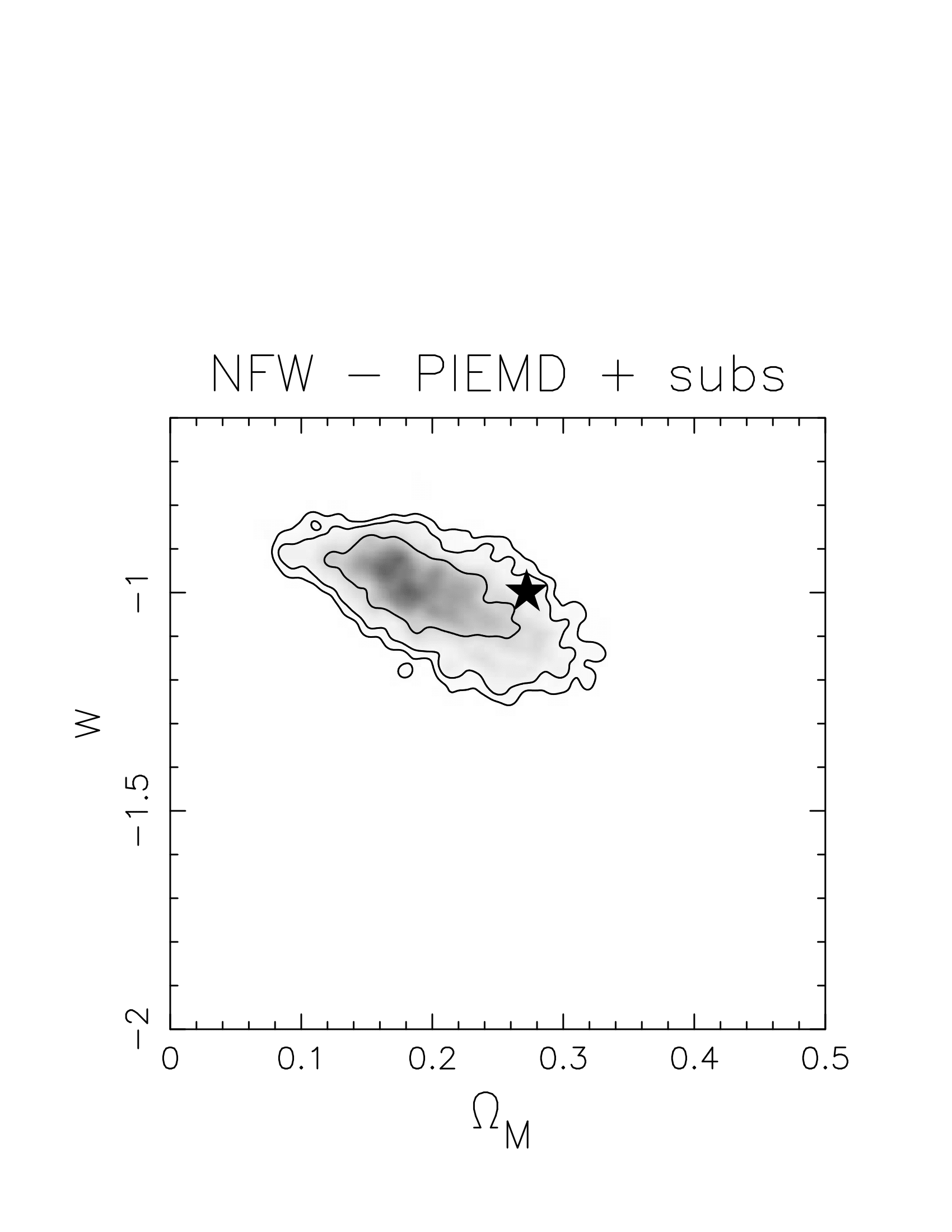}
	\includegraphics[trim=5cm 2cm 2.5cm 5cm, width=4cm]{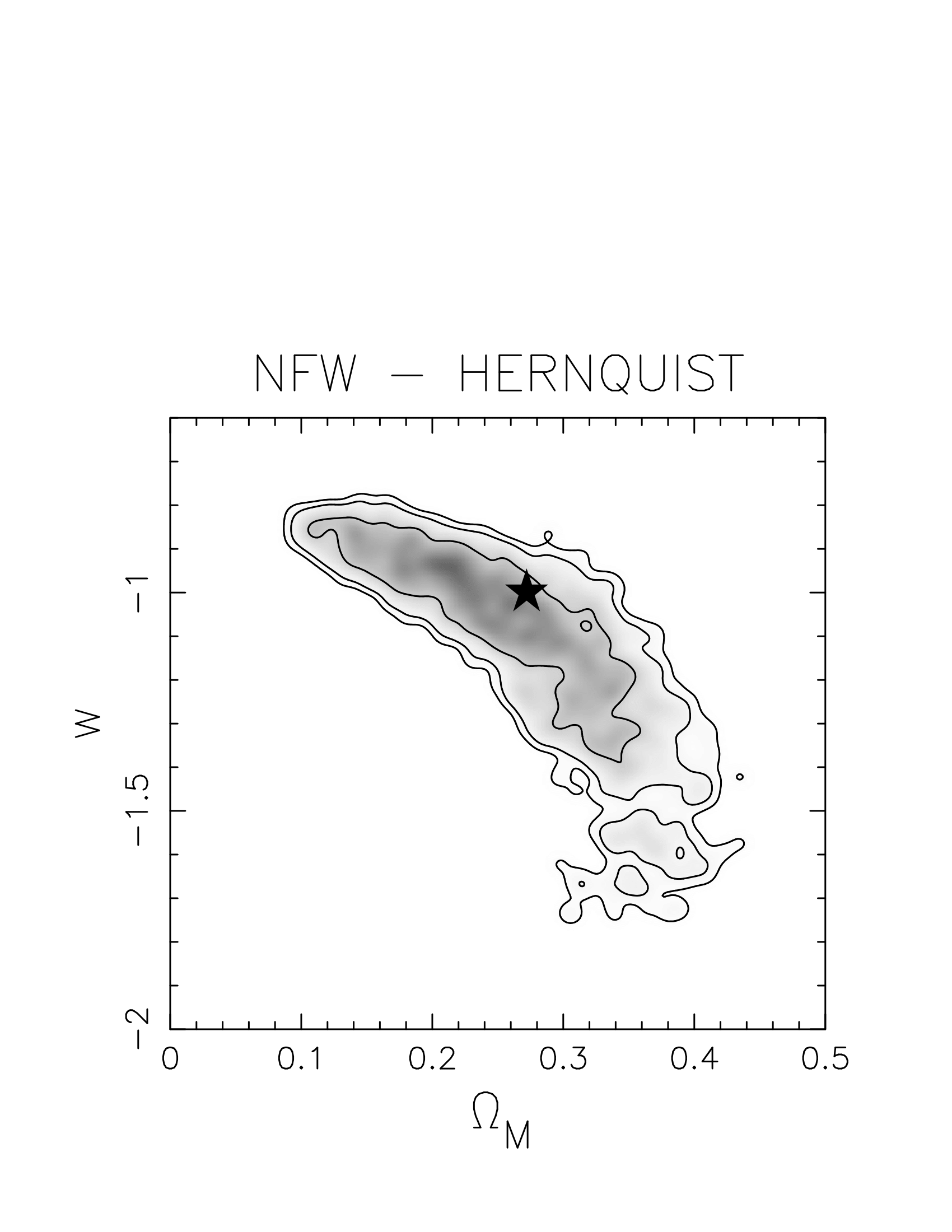}
	\includegraphics[trim=5cm 2cm 2.5cm 4cm, width=4cm]{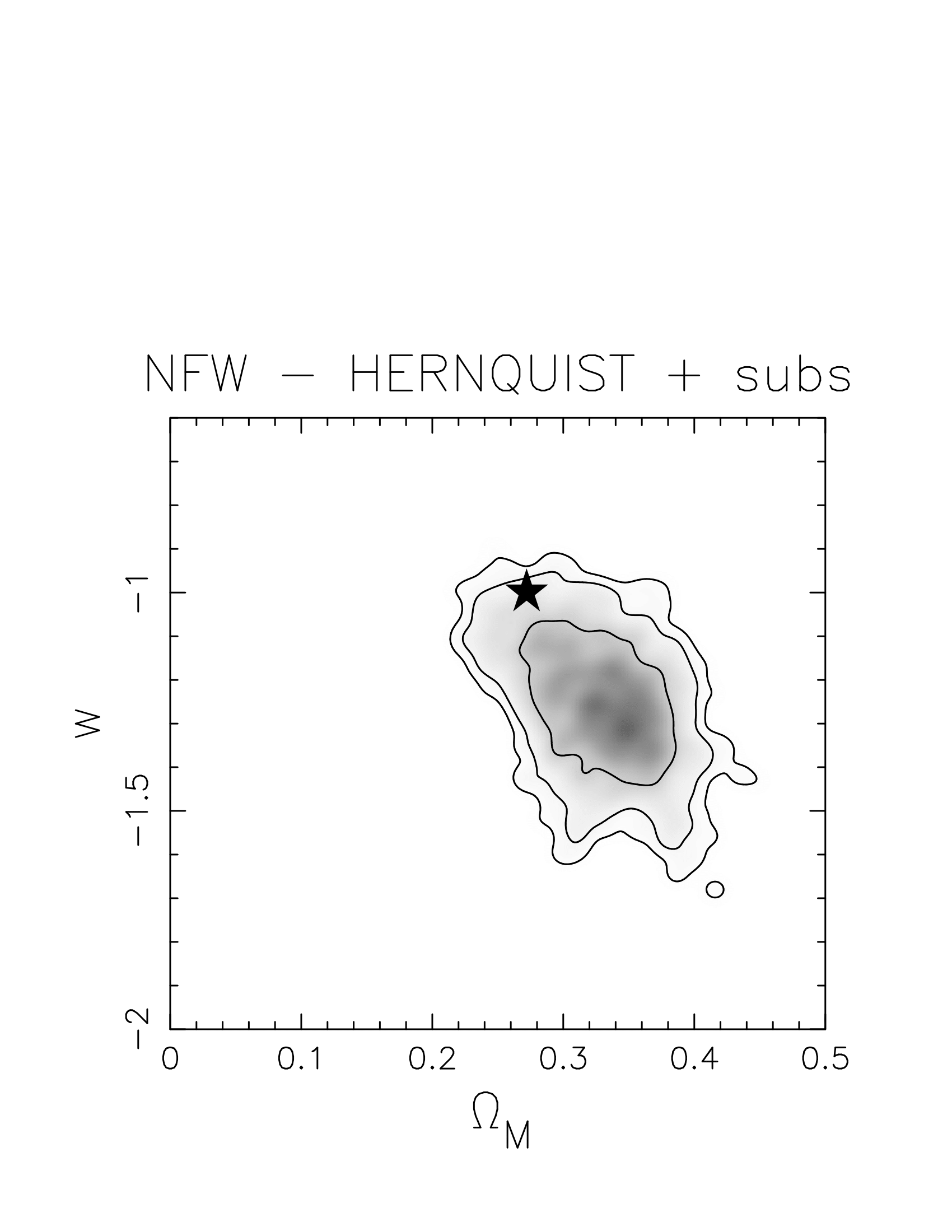}
	\includegraphics[trim=5cm 2cm 2.5cm 4cm, width=4cm]{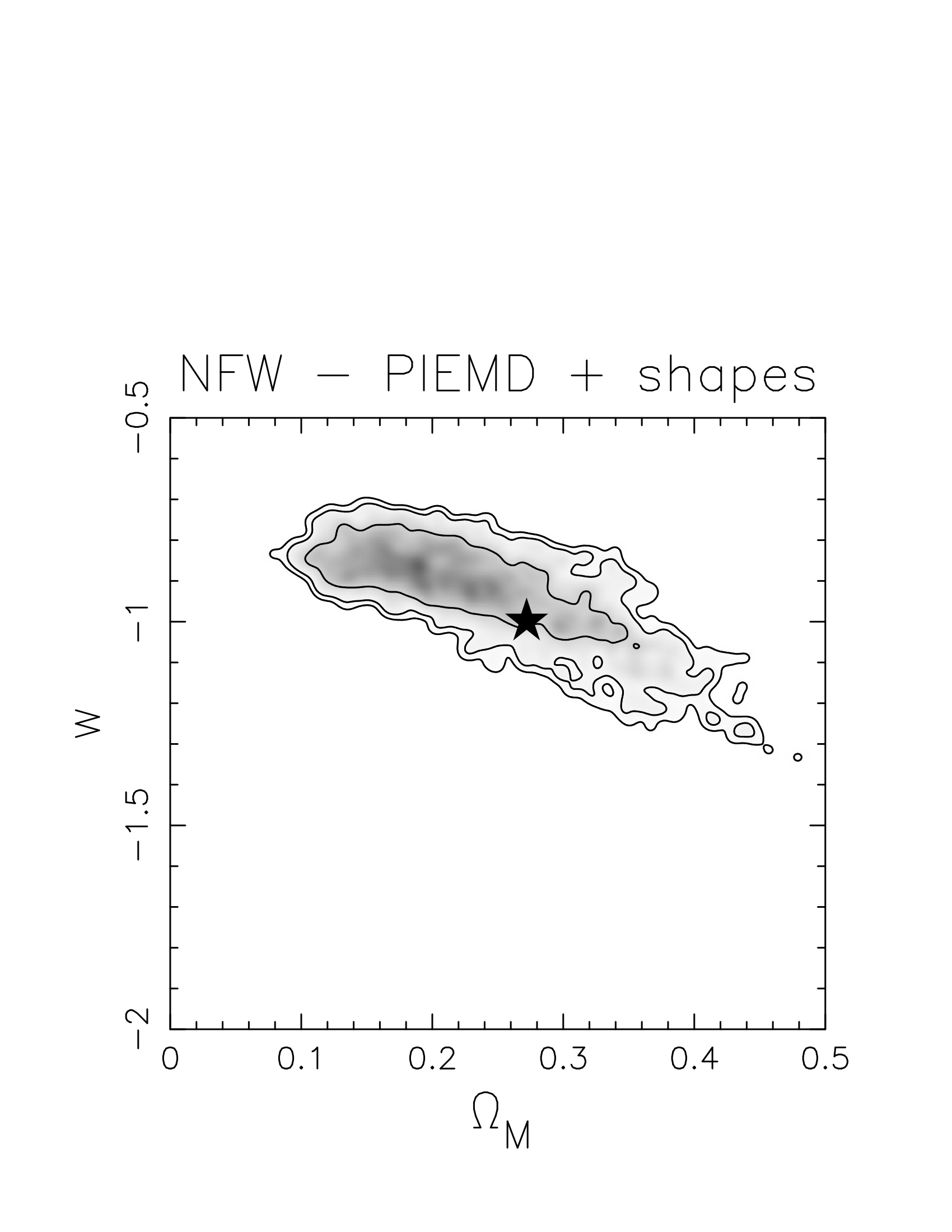}
	\caption{Estimation of cosmological parameters $\mathrm{\Omega_{M}}$-$w$  for the \textit{Ares} cluster using all multiple images. Above each panel the density profile used to fit the smooth component and BCGs is specified.The plotted contours are the 1, 2 and 3-$\sigma$ confidence levels and the star indicates the true value.}\label{cosmotousares}
\end{figure*}

\begin{figure*}
	\includegraphics[trim=5cm 1.5cm 2.5cm 4cm, width=4cm]{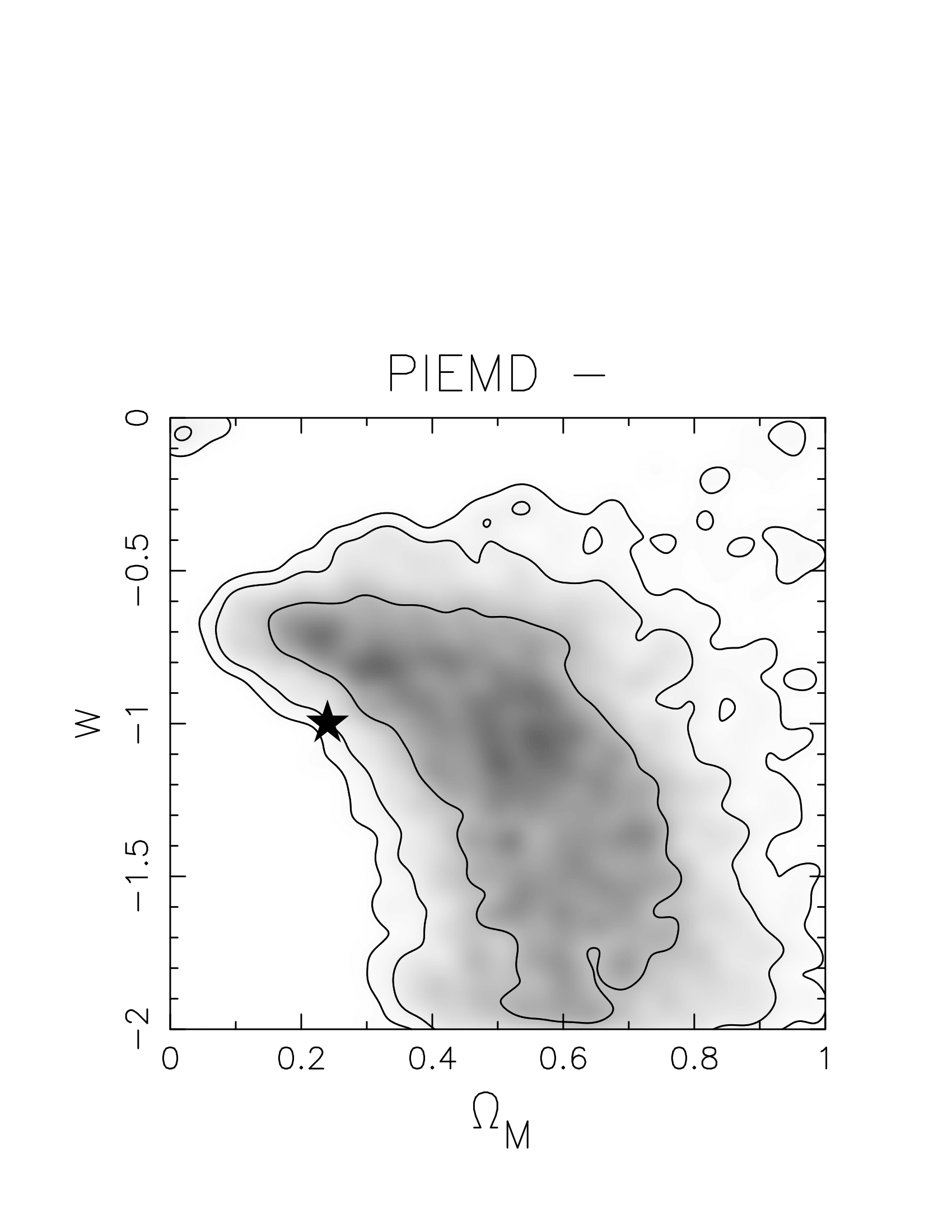}
	\includegraphics[trim=5cm 1.5cm 2.5cm 4cm, width=4cm]{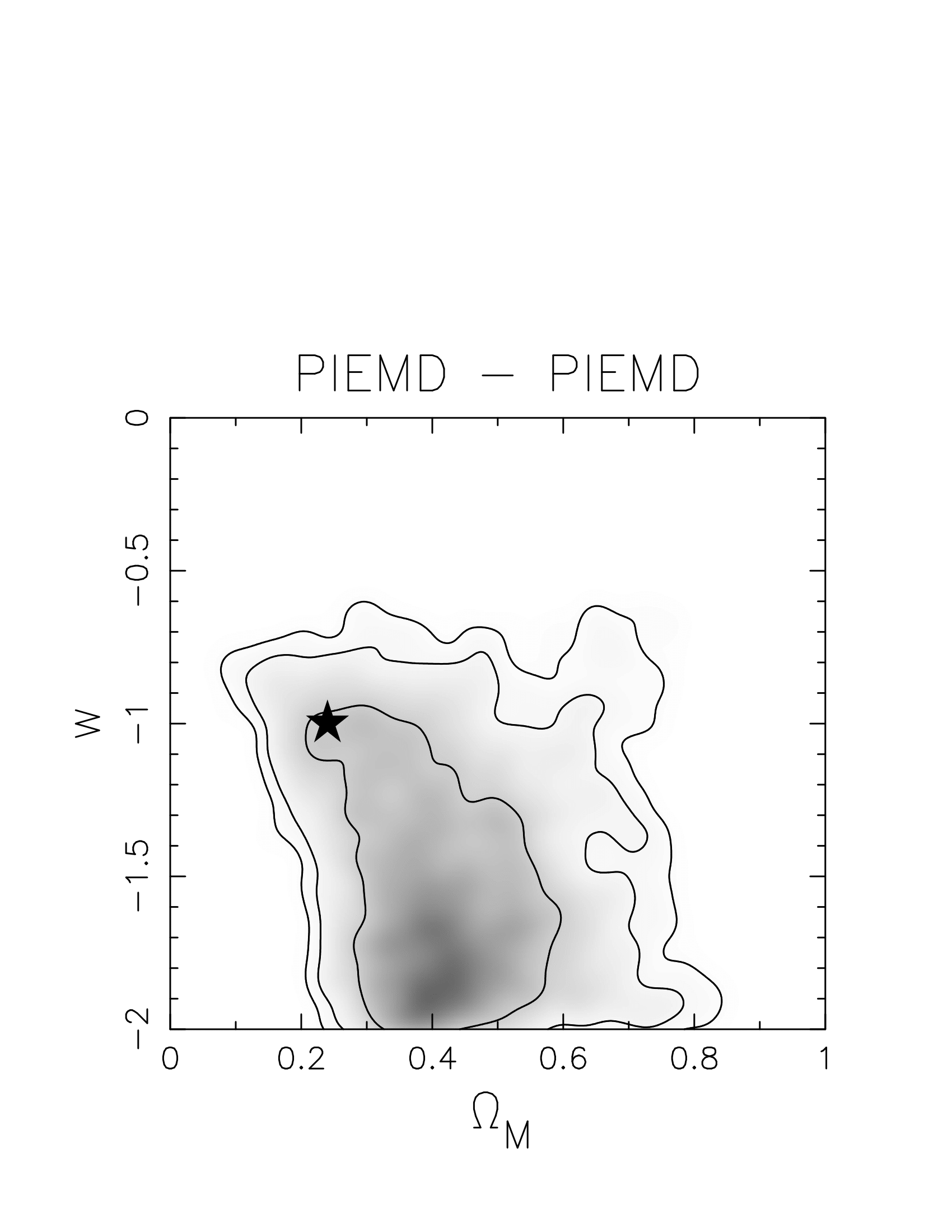}
	\includegraphics[trim=5cm 1.5cm 2.5cm 4cm, width=4cm]{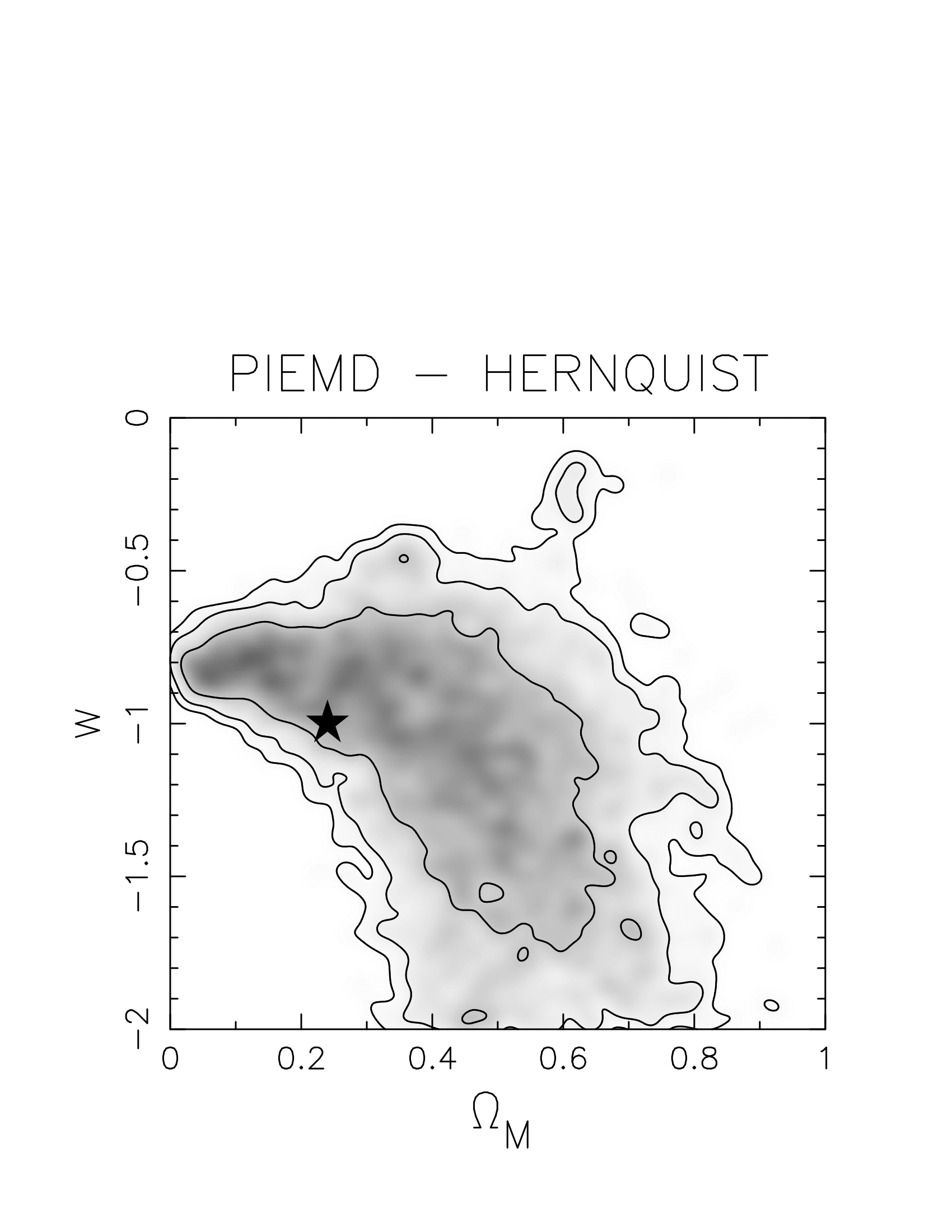}\\
	\includegraphics[trim=5cm 1.5cm 2.5cm 4cm, width=4cm]{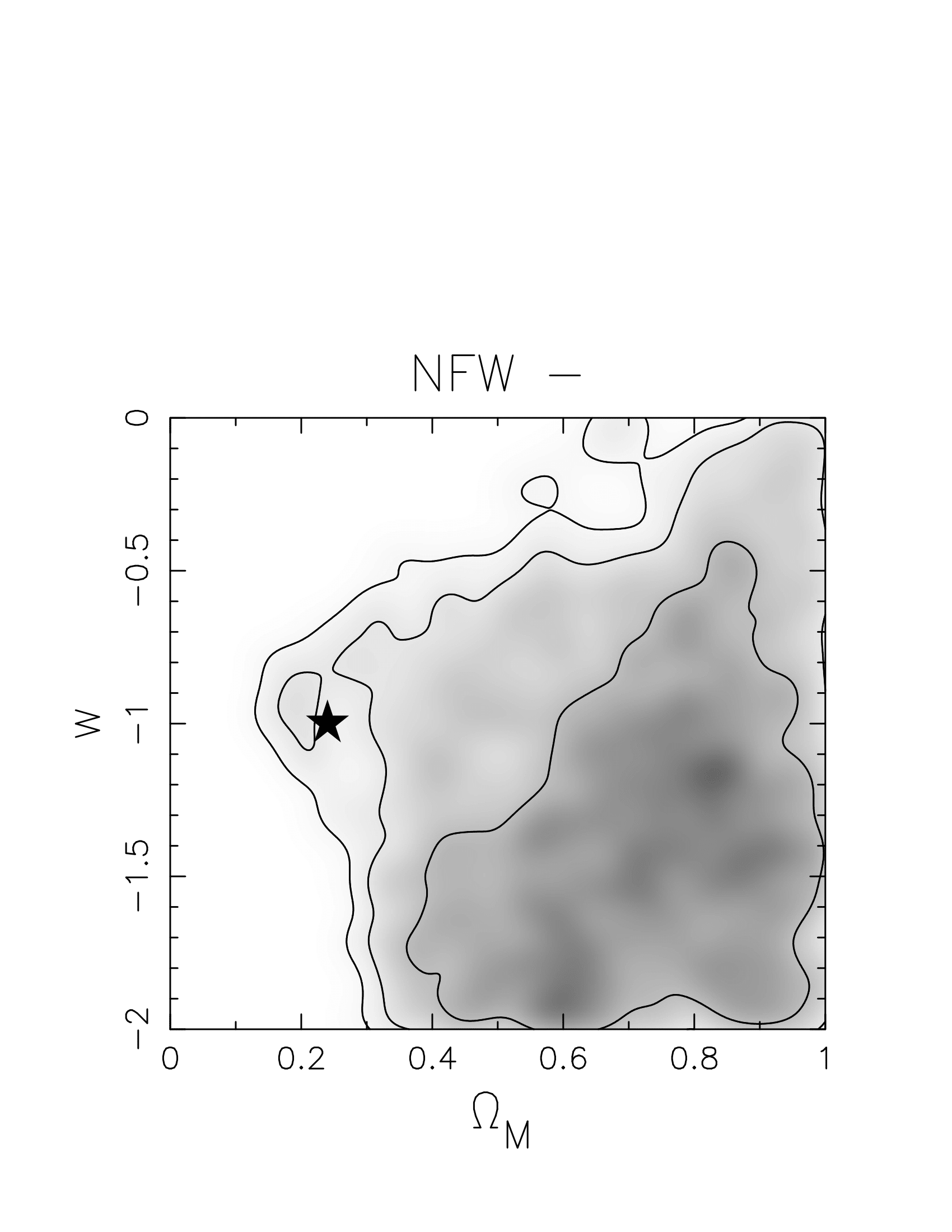}
	\includegraphics[trim=5cm 1.5cm 2.5cm 4cm, width=4cm]{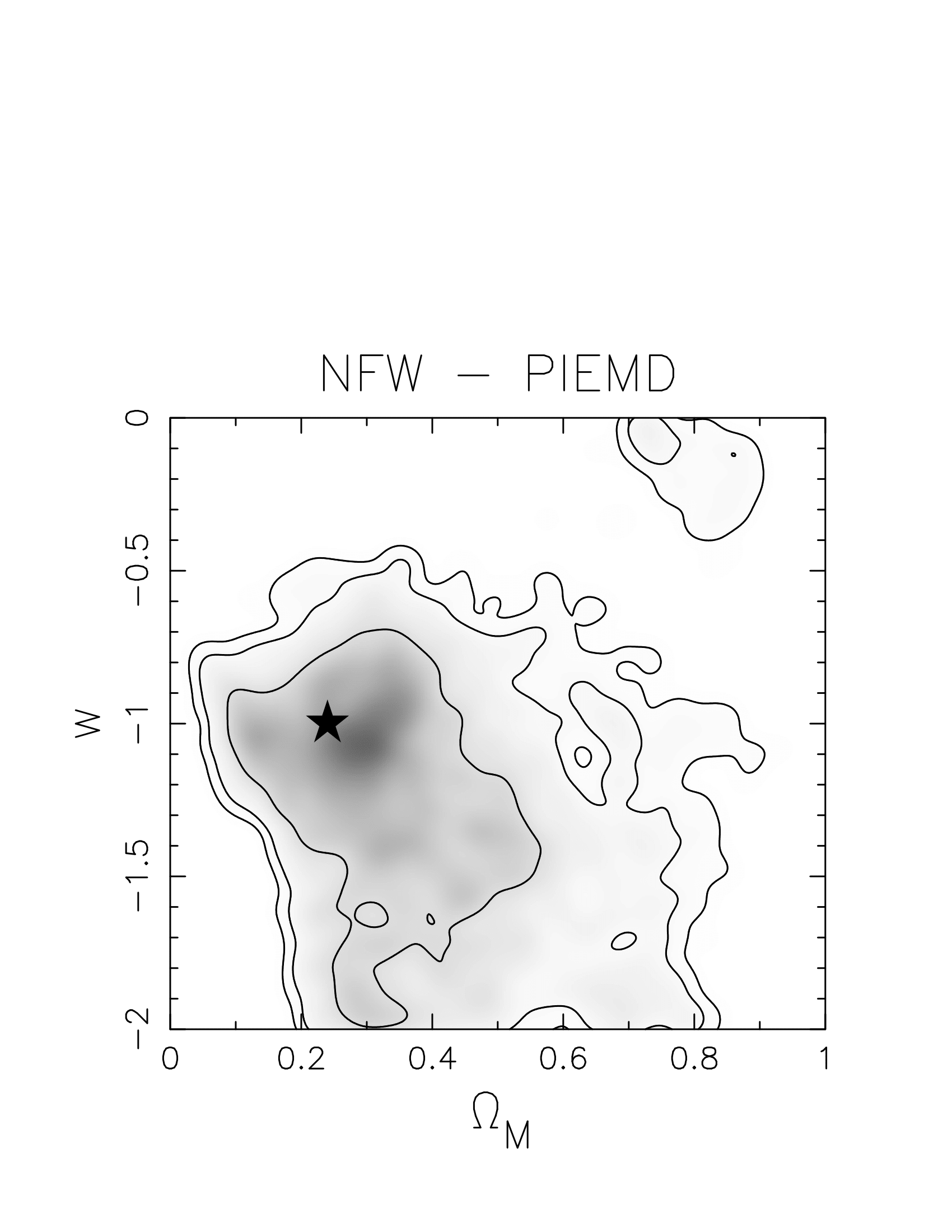}
	\includegraphics[trim=5cm 1.5cm 2.5cm 4cm, width=4cm]{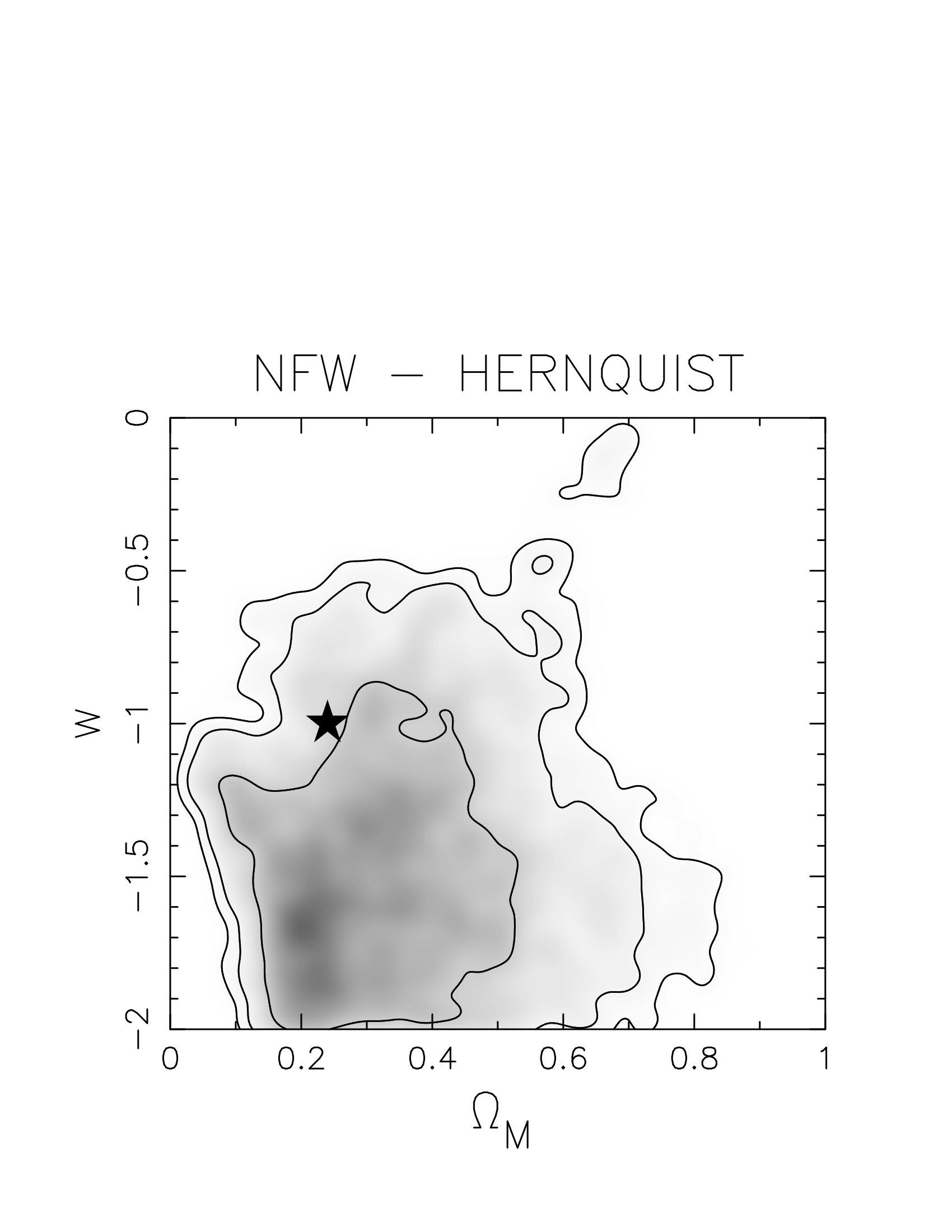}
	\caption{Estimation of cosmological parameters $\mathrm{\Omega_{M}}$-$w$  for the \textit{Hera} cluster using all multiple images. Above each panel the density profile used to fit the smooth component and BCGs is specified. The plotted contours are the 1, 2 and 3-$\sigma$ confidence levels and the star indicates the true value.}\label{cosmotoushera}
\end{figure*}

As mentioned before, for each model we have (taking into account all multiple images with spectroscopic redshifts), the cosmological parameters $\mathrm{\Omega_{M}}$ and $w$ are left as free parameters. 
The constraints obtained are shown in Figure \ref{cosmotousares} for \textit{Ares} and Figure \ref{cosmotoushera} for \textit{Hera}.\\
Using either different density profiles for the large-scale clumps and BCGs provides similar constraints on the $\mathrm{\Omega_{M}}$ and $w$ parameter-space (within the 1$\sigma$ contours) if a large number of multiple images is available (with spectroscopic redshifts and with a positional error of 0.5") and if the model is realistic enough (all but model 1:\textit{ PIEMD},which has the largest mass bias in Figure \ref{biasmass}).\\
However, including in the modelling massive substructures in the cluster's outskirts translates into a decrease of the statistical errors as the cosmological contours are smaller: at the risk of biasing the results. Our work is in agreement with these previous studies showing that massive structures in the outskirts of clusters do impact, not only the mass distribution but also the constraints on cosmological parameters yielding smaller contours. Not only do the line-of-sight structures introduce a systematic error in the strong lensing modelling \citep[eg.][]{Bayliss2014, Giocoli2016} but also distant massive structures in the lens plane have a considerable impact in the position of multiple images \citep{Tu2008, Limousin2010} and thus on the mass constraints. Indeed, in \citet{Mahler2017} the authors evaluate the impact of the presence of mass clumps in the outskirts of the cluster core of Abell 2744 \citep{Jauzac2016} to change the mass profile by $\sim6\%$ at 200kpc. \citet{McCully2017} have shown that considering the environmental and line of sight perturbations should not be taken aside in the modelling as in doing so, the fit does not reproduce the input lens system parameters or the Hubble constant.\\
The contours obtained with \textit{Hera} are, in general, much larger than those from \textit{Ares}. This can be explained by the fact that the redshift range of \textit{Ares}'s sources is twice larger than for \textit{Hera}. By running one of \textit{Ares} models with multiple images only up to $z \sim 3.5$, we checked that, the contours were larger indeed.
The widening of the contours might then not be due to the number of images taken into account as seen in Figure \ref{fam}. The left panels show the constraints on the $\mathrm{\Omega_{M}}$ - $w$ parameter space for one of \textit{Ares} configurations. Considering the same number of multiple images, but spanning a twice larger range of redshift, provides tighter constraints than those obtained with Hera.\\
Finally, we would like to point out that recovering cosmological parameters with strong lenses has been until now performed for unimodal clusters (or more simple clusters than the FF) which would be the preferred configuration for cosmography. Recently, \citet{Caminha2016} performed the first cosmography analysis with a Frontier Fields cluster, AS1063, which is the most relaxed cluster of the sample. By carefully selecting a sub-sample of secured multiple images they achieve a rms of $0.3"$ and put constraints on the $\mathrm{\Omega_{M}}$, $w$ and $\Omega_{\Lambda}$ parameters. \\ 
However, we show in this work that complex and multimodal clusters such as \textit{Ares} can yield tight and competitive constraints on the $\mathrm{\Omega_{M}}$ - $w$ parameter space.

%%%%%%%%%%%%%%%%%%%%%%%%%%%%%%%%%%%%%%%%%%%%%%%%

%%%%%%%%%%%%%%%%%%%%%%%%%%%%%%%%%%%%%%%%%%%%%%%%%%
\subsection{Redshift catalogues} \label{redshifts}
In this section we investigate how the estimation of robust cosmological parameters is affected by the availability of different sets of constraints. We extend this study for three of the models for \textit{Ares} in Section \ref{[section4]} (models 2, 3, 6). \\
These analyses have been only carried out for the \textit{Ares} cluster as it has three times more multiple images (see Table \ref{table:1}) and the range of redshift is twice wider than for \textit{Hera}.

\subsubsection{Redshift range} \label{zrange}
\begin{figure*}
		\centering
		\includegraphics[width=\linewidth]{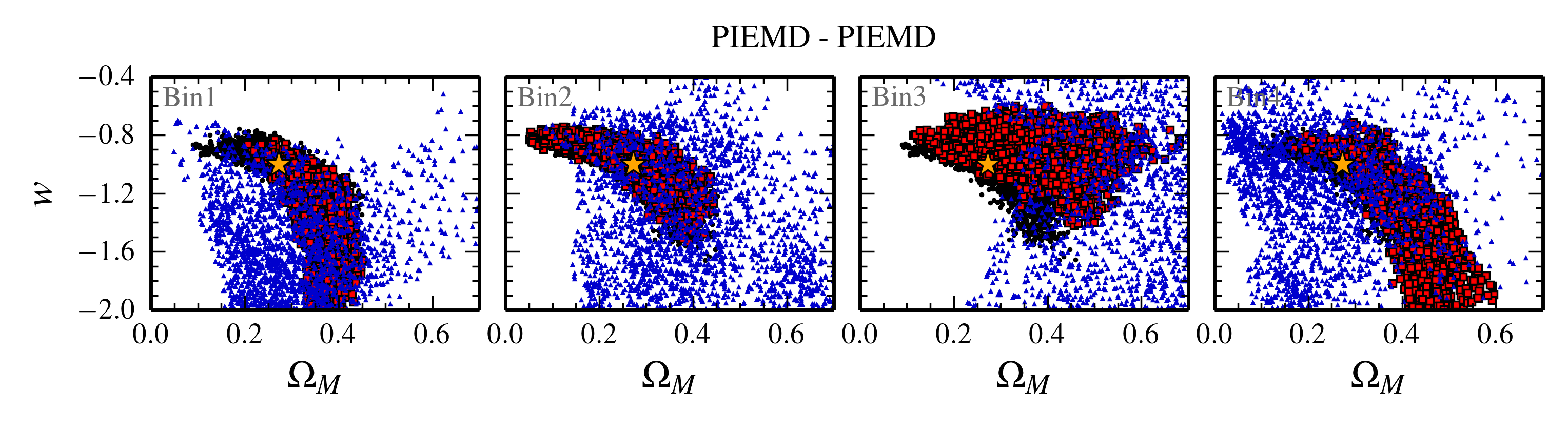} 
		\includegraphics[width=\linewidth]{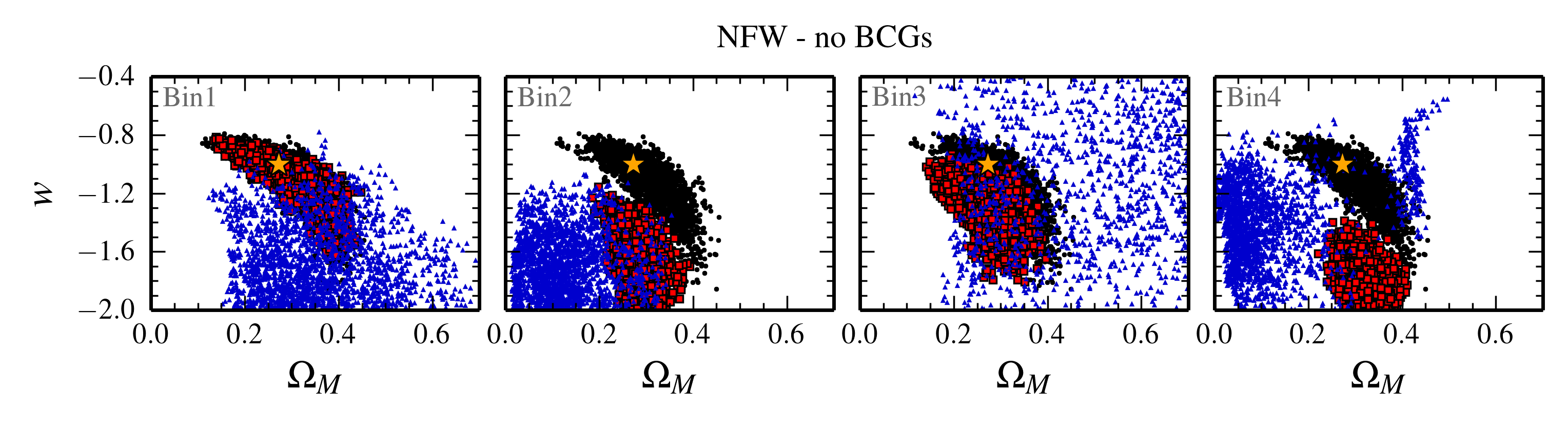} 
		\includegraphics[width=\linewidth]{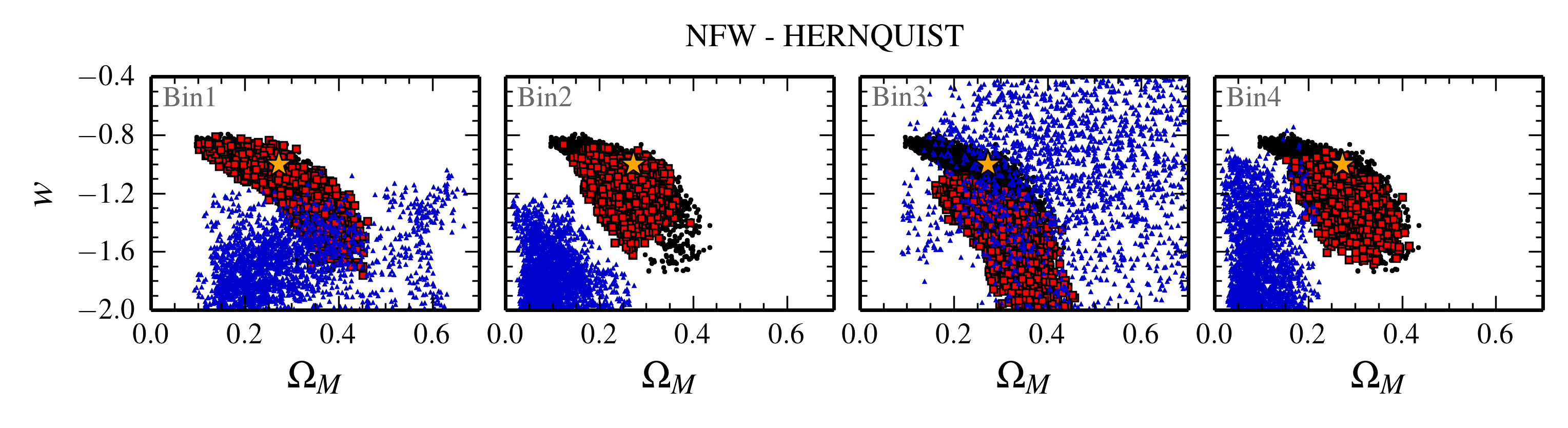} 
		\caption{Constraints on the $\mathrm{\Omega_{M}}$ and $w$ parameters for \textit{Ares}. As filled black circles, the fiducial constraints (all multiple images with spectroscopic redshifts) presented in Figure \ref{cosmotousares}. When considering only arcs within the redshift range specified in each panel with spectroscopic information (blue filled triangles) the constraints are obtained biased. If we consider, not only arcs within the redshift range specified in each panel with spectroscopic information, but also the remaining catalogue with photometric redshifts (with a $0.04(1+z)$ precision) there is no bias on the cosmological constraints (shown by the filled red squares) or at least, it is reduced.}\label{binz}
	\end{figure*}

We investigate first whether there is a redshift range of background sources more efficient to recover the input cosmology and if photometric redshifts can complete our samples.\\
We split the redshift catalogue into four bins of redshift: \\
\vspace{0.03cm}\\
$\bullet$ \textit{Bin1} as $z_{s} \leq 1.96$, \\
$\bullet$  \textit{Bin2} as $1.96 < z_{s} \leq 3.08$,\\ 
$\bullet$  \textit{Bin3} as $3.08 < z_{s} \leq 3.68$,\\
$\bullet$  \textit{Bin4} as $ z_{s} > 3.68$.\\
\vspace{0.03cm}\\
and we build the new multiple images catalogues as follows:\\
for each of the new four multiple images catalogue we keep the images whose redshift is inside each bin as spectroscopic and the rest as photometric redshifts which have been virtually created with a precision of $0.04(1+z)$ and assuming there are no catastrophic errors. This precision is already achievable with the HFF data: \citet{Castellano2016} determined a typical photometric redshift error on the multiple images in Abell 2744 and MACS 0416 of $3-5\%$.
Of the 242 total images provided we keep $\sim$ 60 with spectroscopic information per bin and the rest with photometric accuracy.\\
We also study if it is preferable to only use spectroscopic information or add photometric redshifts as constraints so we also use multiple images catalogues with only images belonging to each bin (having then $\sim$ 60 images, with spectroscopy).
We compare, for each bin of redshift stated above, the constraints on the $\mathrm{\Omega_{M}}$-$w$ parameter-space with photometric redshifts as additional information or not.\\
The results are shown in Figure \ref{binz}. For reference, the fiducial models of Figure \ref{cosmotousares} are also shown (black constraints). First, we show that considering only spectroscopic redshifts from a certain redshift range (blue points) biases the estimation of cosmological parameters whatever the redshift bin considered is, the bias being similar whichever the profile and/or configuration used are for the same bin considered.\\
By completing these spectroscopic redshifts catalogues with multiple images with photometric information (red points), spanning the ranges in redshift not covered by them, we recover unbiased cosmology (golden star) in most cases or at least reduce this bias. Note that, if cosmological parameters are still biased, it is the equation of state parameter $w$ which is more affected than $\Omega_M$.\\
Photometric redshifts are then a useful piece of additional information to take into account in the modelling and completing the spectroscopic catalogue if the latter covers only a narrow redshift range.

\subsubsection{Photometric families} \label{zphot_fam}
\begin{figure*}
	\centering
	\includegraphics[width=\linewidth]{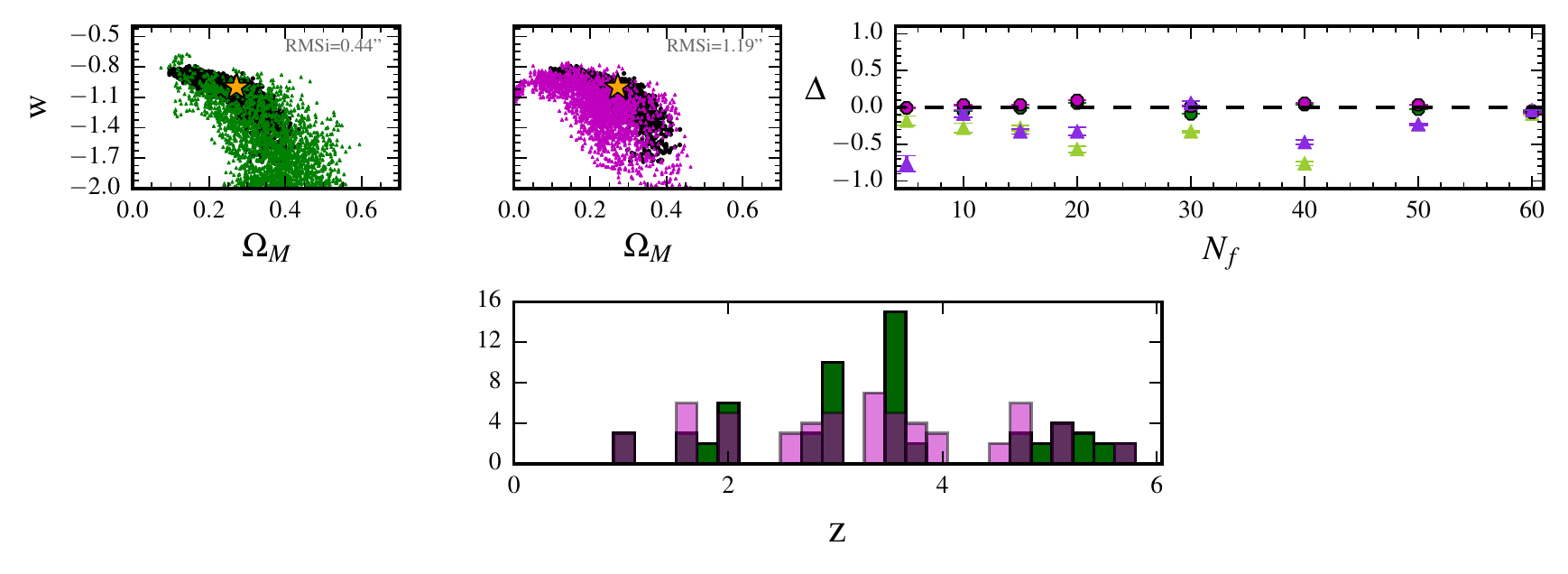} 
	\caption{Top left and middle panels: constraints on the $\mathrm{\Omega_{M}}$ and $w$ parameters for \textit{Ares} for the model NFW - HERNQUIST. As filled black circles, the fiducial constraints (all multiple images with spectroscopic redshifts); as green and magenta filled triangles are plotted the constraints from two models with a (slightly different) reduced catalogue of spectroscopic redshifts (shown in the bottom panel). Right panel: systematic bias in the recovery of cosmological parameters as a function of increasing photometric families added in the modelling (with a $0.04(1+z)$ precision). The filled triangles show the bias on the $w$ parameter and the filled circles, on the $\Omega_M$ parameter. The statistical error is represented by the error bars. }\label{fam}
\end{figure*}

Considering multiple images from a restricted range of redshift leads to an estimation of wider constraints (which are biased) on the $\mathrm{\Omega_{M}}$ and $w$ parameter space whatever the mass distribution of the cluster is. This bias is due to the narrow range of redshifts considered and not to a reduced number of multiple images considered as seen below.\\
In this section, we consider as constraints a reduced catalogue of multiple images with spectroscopic redshifts by creating two similar catalogues (but with a different redshift distribution) which contain $\sim$ 60 images and spanning all the redshift range available (see histograms in Figure \ref{fam}). We have considered the same three mass models (id 2, 3 and 6) of Section \ref{[section4]} for \textit{Ares}. For clarity, we only show the results obtained for one of them in Figure \ref{fam} as the results were very similar for the 3 of them. \\
In the left and middle panels we show in black the constraints obtained with the fiducial model NFW- HERNQUIST of (Section \ref{cosmotousares}), on top are the constraints with the two reduced catalogues. As we can notice, the cosmological parameters are unbiased.\\
We investigate if adding an increasing number of families of multiple images with photometric redshifts (spanning all the redshift range) in the modelling translates into an improvement in the cosmology estimation (i.e a smaller systematic than statistical error). This is shown in the right panels of Figure \ref{fam} where the systematic and statistical errors (errors bars) are plotted as a function of the increasing number of photometric families taken in account. The coloured circles show the bias on $\Omega_M$, the triangles on $w$. We observe the same behaviour for the three models, where $\Omega_M$ is less affected by than $w$ the modelling and the statistical errors are smaller than the systematic uncertainties. $w$ is systematically underestimated. We do not report, however, any trend with the increasing number of photometric families. 

\subsection{On the positional uncertainty of the multiple images} \label{[sigmapos]}
\begin{figure*}
	\centering
	\includegraphics[width=\linewidth]{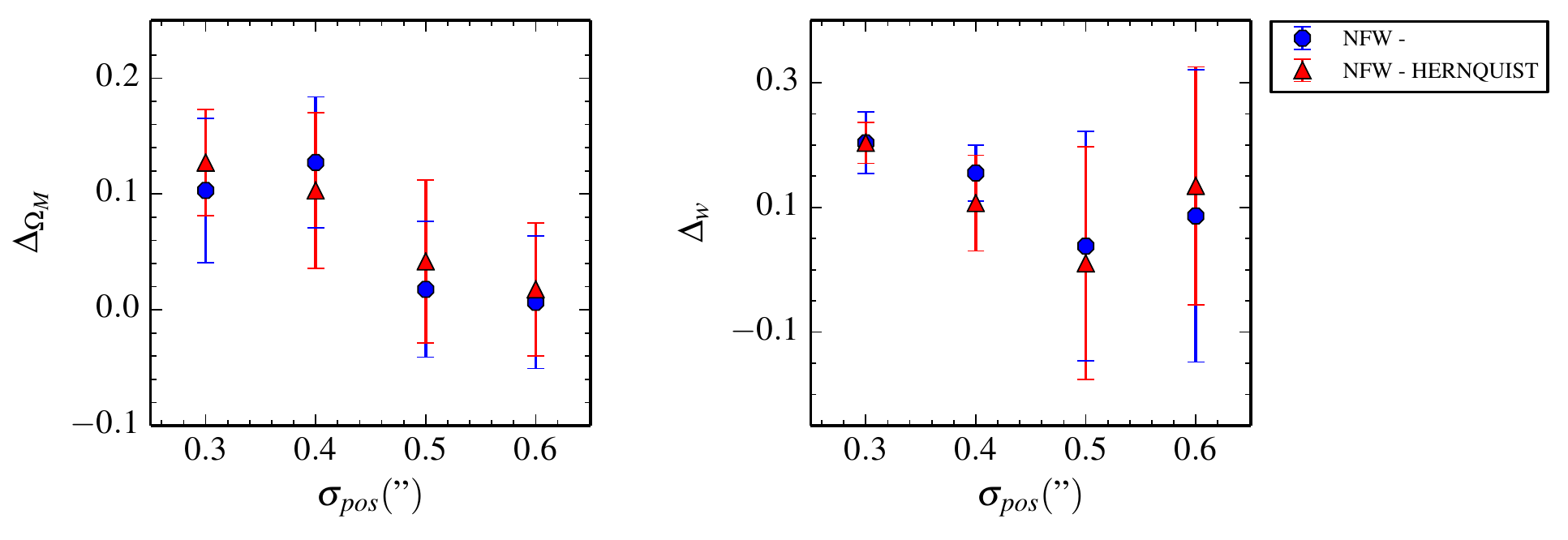} 
	\caption{Bias on the estimation of $\mathrm{\Omega_{M}}$ (left panel) and $w$ (right panel) depending on the assumed positional error of the multiple images for two mass distributions for \textit{Ares} (the NFW - HERNQUIST being more accurate)}
	\label{biaserrpos}
    \end{figure*}

Throughout this paper, we have assumed an uncertainty of ~0.5" for the position of multiple images, closer value to the RMS in the image plane thus providing a reduced $\chi^{2}\sim1$. We show in Figure \ref{biaserrpos} the bias on the estimation of $\mathrm{\Omega_{M}}$ and $w$ for three positional errors of the multiple images assumed in the modelling for two models of Section \ref{[section4]} for \textit{Ares}. This bias is the lowest for a positional uncertainty of the order of the RMS.  However, this Figure also shows that underestimating the uncertainties on the observations can lead to biased constraints on the $\mathrm{\Omega_{M}}$-$w$ parameter space.

\section{Conclusions} \label{[conclu]}
We have analysed two mock galaxy clusters, \textit{Ares} and \textit{Hera} from the FF-SIMS Challenge \citep{meneghetti2016}, both complex and bi-modal, comparable to the FF clusters. We have investigated the systematic errors in the strong lensing parametric modelling arising from the choice of the density profiles and configurations as well as from the availability of constraints (spectroscopic or photometric redshifts) and therefore the impact on the retrieval of robust cosmological parameters. \\
Our main conclusions are the following:
\begin{enumerate}
	\item Galaxy clusters are not isolated systems and can present large structures in the cluster outskirts \citep{Jauzac2016, Foex2017}. With this work we provide further evidence that distant massive substructures in the lens-plane of galaxy clusters have a significant impact on the mass distribution. Wide-field imaging around massive clusters is thus needed to account for these structures in the modelling. In an era of precise cosmology, we show that the cluster's environment cannot be ignored in order to yield a more precise mass reconstruction and therefore competitive constraints on $\mathrm{\Omega_{M}}$ and $w$.\\
	\item As expected, the smaller the bias on the mass, the smaller the bias on cosmological parameters. The bias on the total mass is a quality indicator for the cosmological constraints. On the other hand, magnification, the cluster's ellipticity and orientation do not allow to discriminate between models (assuming the same modelling technique).  \\	
	\item Considering a positional error of $0.5"$, the estimation of cosmological parameters is not affected by the choice of different mass profiles or configurations when a sufficient number of constraints is available ($n_{im}> \sim 60$).\\
	\item The bias on the estimation of cosmological parameters is the lowest for a positional uncertainty of the order of the RMS. Underestimating the uncertainty on the observations can lead to biased constraints on the $\mathrm{\Omega_{M}}$-$w$ parameter space. \\		
	\item Considering multiple images, from a restricted range of redshift leads to an estimation of biased cosmological parameters. Taking into account multiple images from a broader range of redshift with photometry information can correct this bias or, at least, reduce it.\\			
	\item We do not report any trend between an increasing number of photometric families taken into account in the modelling and a more precise estimation of $\mathrm{\Omega_{M}}$ and $w$. \\
	\item $\mathrm{\Omega_{M}}$ is less sensitive to systematic errors than $w$ this latter being systematically underestimated when recovered biased. 
\end{enumerate}
Stronger constraints can be obtained by combining the estimates on $\mathrm{\Omega_{M}}$ and $w$ from several strong lensing clusters \citep{Daloisio2011}. We show that, not only unimodal \citep[or simpler clusters than \textit{Ares} and \textit{Hera};][]{jullo2010, Caminha2016}, but also more complex and multimodal clusters can yield competitive constraints. Upcoming surveys such as \textit{James Webb Space Telescope} will make strong lensing cosmography a very powerful tool by detecting an even larger number of arcs than currently with HST. 

\section*{Acknowledgments}
We thank the referee for her/his useful comments that helped improving this work. 
A.A is grateful to Gabriel Bartosch Caminha for useful and constructive discussions. We also thank M. Meneghetti,  P. Natarajan and D. Coe for providing the \textit{Ares} and \textit{Hera} mock clusters for the Frontier Fields Lens Modeling Comparison Project. \\
This work has been carried out thanks to the support of the OCEVU Labex (ANR-11-LABX-0060) and the A*MIDEX project (ANR-11-IDEX-0001-02) funded by the "Investissements d'Avenir" French government program managed by the ANR. \\
This work was granted access to the HPC resources of Aix-Marseille Universit\'{e} financed by the project Equip@Meso (ANR-10-EQPX-29-01) of the program "Investissements d'Avenir" supervised by the Agence Nationale pour la Recherche. \\
We are also grateful to CNES for financial support. M. L thanks CNRS for financial support.\\
CG acknowledges support from the Italian Ministry for Education, University and Research (MIUR) through the SIR individual grant SIMCODE, project number RBSI14P4IH\\
M.J acknowledges support by the Science and Technology Facilities Council [grant number ST/L00075X/1].

%%%%%%%%%%%%%%%%%%%%%%%%%%%%%%%%%%%%%%%%%%%%%%%%%%



\bibliographystyle{mnras}
\bibliography{biblio3} % if your bibtex file is called example.bib

%%%%%%%%%%%%%%%%%%%%%%%%%%%%%%%%%%%%%%%%%%%%%%%%%%

%%%%%%%%%%%%%%%%% APPENDICES %%%%%%%%%%%%%%%%%%%%%

\appendix

\section{Source redshift distribution}

\begin{figure}
		\centering
		\includegraphics[width=0.9\columnwidth]{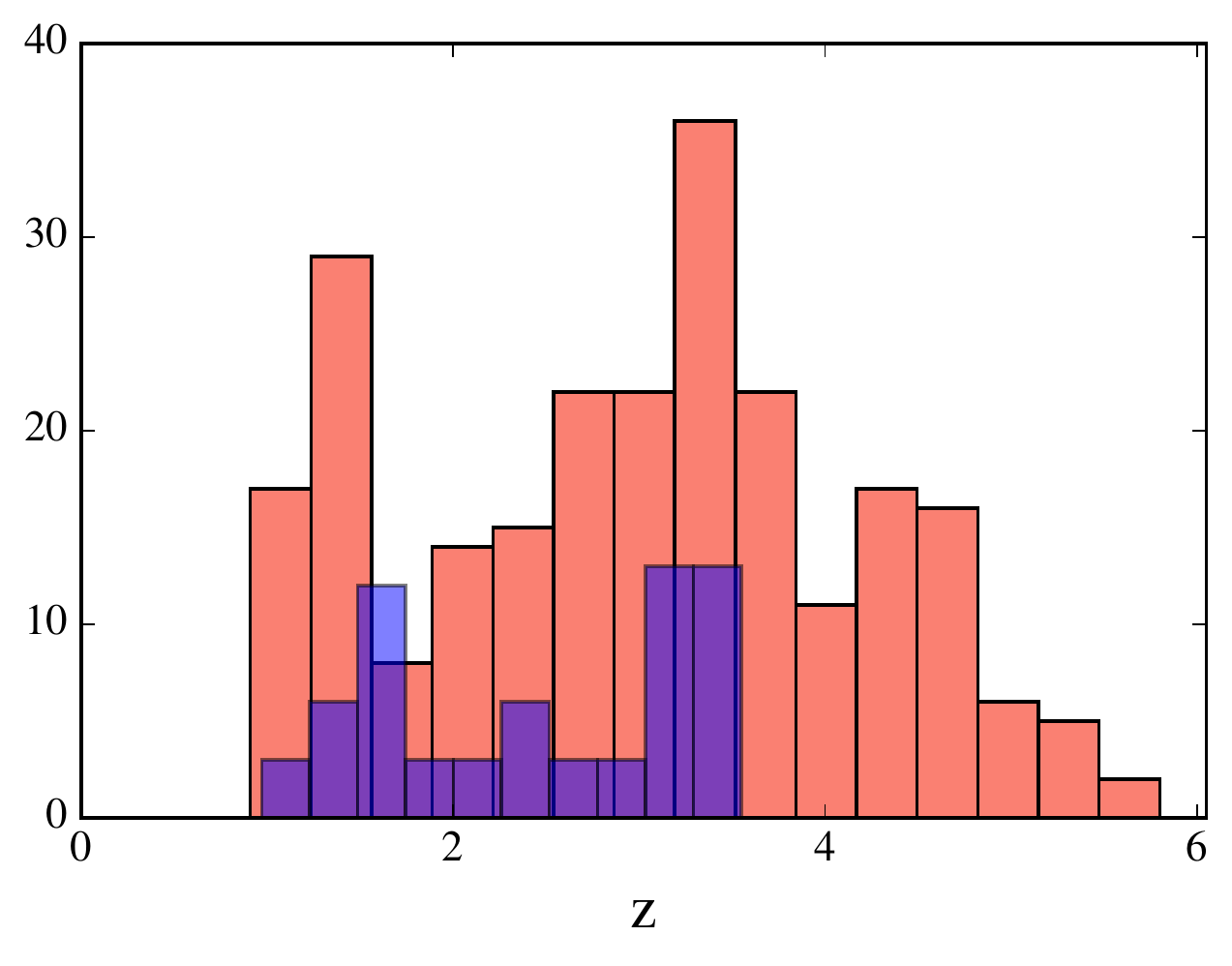} 
		\caption{Resdhift distribution of the background sources for the strong lens \textit{Ares} in pink and for \textit{Hera} in blue.}
		\label{z_areshera}
	\end{figure}

We show in Figure \ref{z_areshera} the redshift distribution of background sources for the \textit{Ares} and \textit{Hera} clusters

%%%%%%%%%%%%%%%%%%%%%%%%%%%%%%%%%%%%%%%%%%%%%%%%%%

% Don't change these lines
\bsp	% typesetting comment
\label{lastpage}
\end{document}